\documentclass[]{aa}  
\usepackage{array, multirow, graphicx}
\usepackage[dvipsnames]{xcolor}
\usepackage{wasysym}[integrals]
\usepackage{amssymb}
\usepackage{pifont}
\usepackage{siunitx}
\usepackage{booktabs, tabularx}

\usepackage{txfonts}
\usepackage[]{hyperref}
\hypersetup{colorlinks,citecolor=blue, linkcolor=blue, urlcolor=blue, breaklinks=true}
\newcounter{magicrownumbers}

\usepackage{cellspace, makecell} 
\usepackage{mathtools}

\newcommand{\tonda}[1]{\!\left ( #1 \right )}
\newcommand{\quadra}[1]{\left [ #1 \right ]}
\newcommand{\abs}[1]{\left | #1 \right |}
\newcommand{\sopra}[1]{\overline{#1}}
\newcommand{\eq}[1]{\begin{equation} #1 \end{equation}}
\newcommand{\cii}{[C\RNum{2}]$_{\rm 158\,\mu m}$}
\newcommand{\RNum}[1]{\uppercase\expandafter{\romannumeral #1\relax}}

\begin{document} 

\newcommand{\red}[1]{\textcolor{red}{#1}}
\newcommand{\green}[1]{\textcolor{green}{#1}}

   \title{ALMA survey of a massive node of the Cosmic Web at $z\sim 3$}
        \subtitle{II. A dynamically cold and massive disk galaxy in the proximity of a hyperluminous quasar}

   \author{A.~Pensabene
          \inst{\ref{unimib}}
         	\and
	S.~Cantalupo
	\inst{\ref{unimib}}
		\and
	W.~Wang
	\inst{\ref{unimib}}
		\and
	C.~Bacchini
	\inst{\ref{dark}}%, \ref{inaf-pd}}
        		\and
        	F.~Fraternali
	\inst{\ref{kapt}}
		\and
	M.~Bischetti
	\inst{\ref{units}, \ref{inaf-oats}}
		\and
	C.~Cicone
	\inst{\ref{ITA}}
		\and
	R.~Decarli
        \inst{\ref{inaf-bo}}
		\and
	G.~Pezzulli
	\inst{\ref{kapt}}
		\and
         M.~Galbiati
	\inst{\ref{unimib}}
		\and
	T.~Lazeyras
	\inst{\ref{unimib}}
		\and
	N.~Ledos
	\inst{\ref{unimib}}
		\and
	G.~Quadri
	\inst{\ref{unimib}}
		\and
	A.~Travascio
	\inst{\ref{unimib}}
	}	
	
\institute{Dipartimento di Fisica ``G. Occhialini'', Universit\`a degli Studi di Milano-Bicocca, Piazza della Scienza 3, I-20126, Milano, Italy\\\email{antonio.pensabene@unimib.it}\label{unimib}
   		\and
		DARK, Niels Bohr Institute, University of Copenhagen, Jagtvej 155, 2200 Copenhagen, Denmark\label{dark}
		\and
		Kapteyn Astronomical Institute, University of Groningen, Landleven 12, NL-9747 AD Groningen, the Netherlands\label{kapt}
		 \and
		 Dipartimento di Fisica, Universit\`a di Trieste, Sezione di Astronomia, Via G.B. Tiepolo 11, I-34131 Trieste, Italy\label{units}
		 \and
		 INAF--Osservatorio Astronomico di Trieste, Via G. B. Tiepolo 11, I-34131 Trieste, Italy\label{inaf-oats}
		 \and
		Institute of Theoretical Astrophysics, University of Oslo, P.O. Box 1029, Blindern 0315, Oslo, Norway\label{ITA}
		\and
		INAF--Osservatorio di Astrofisica e Scienza dello Spazio, Via Gobetti 93/3, I-40129 Bologna, Italy\label{inaf-bo}
		}

%   \date{Received XXX accepted YYY}

% \abstract{}{}{}{}{} 
% 5 {} token are mandatory
 
\abstract{Advancing our understanding of the formation and evolution of early massive galaxies and black holes requires detailed studies of dense structures in the high-redshift Universe. In this work, we present high-angular resolution ($\simeq0\rlap{.}{\arcsec}3$) ALMA observations targeting the CO(4--3) line and the underlying 3-mm dust continuum toward the Cosmic Web node MQN01, a region identified through deep multiwavelength surveys as one of the densest concentrations of galaxies and AGN at cosmic noon. At the center of this structure, we identify a massive, rotationally supported disk galaxy {\rm located approximately at $\sim10\,{\rm kpc}$ projected-distance and $\sim-300\,{\rm km\,s^{-1}}$ from a hyperluminous quasar at $z=3.2510$}. By accurately modeling the cold gas kinematics, we determine a galaxy dynamical mass of $2.5\times10^{11}\,{M_{\astrosun}}$ within the inner $\simeq 4\,{\rm kpc}$, and a high degree of rotational support of $V_{\rm rot}/\sigma \approx 11$. This makes it the first quasar companion galaxy confirmed as a massive, dynamically cold rotating disk at such an early cosmic epoch. Despite the small projected separation from the quasar host, we find no clear evidence of strong tidal interactions affecting the galaxy disk. This might suggest that the quasar is a satellite galaxy in the early stages of a merger. Furthermore, our spectroscopic analysis reveals a broad, blueshifted component in the CO(4--3) line profile of the quasar host, which may trace a powerful molecular outflow or kinematic disturbances induced by its interaction with the massive companion galaxy. Our findings show that rotationally supported cold disks are able to survive even in high-density environments of the early Universe.
}

               
   \titlerunning{The ALMA view of MQN01 field}
   \authorrunning{Pensabene et al.}
   \maketitle
   
%
%-------------------------------------------------------------------

\section{Introduction}
\label{sect:introduction} 
Luminous quasars (or QSOs) at high redshift ($z>2$) are among the most powerful sources in the early Universe ($< 3\,{\rm Gyr}$). They are sustained by rapid gas accretion onto a central supermassive black hole (BH; $M_{\rm BH}\apprge10^{8}\,M_{\astrosun}$), accompanied by intense star formation episodes in their host galaxies (${\rm SFR}\apprge 100\,M_{\astrosun}\,{\rm yr^{-1}}$; see, e.g., \citealt{Pitchford+2016, Harris+2016, Duras+2017}). {\rm These properties point toward high-redshift quasars residing} in massive structures of the dark matter (DM) density distribution, often associated with galaxy overdensities -- though results in the literature remain mixed \citep[see, e.g.,][]{Garcia-Vergara+2017, Uchiyama+2018, Arrigoni-Battaia+2023}. These environments promote efficient gas accretion from the Cosmic Web, alongside galaxy interactions, mergers, and feedback processes, ultimately driving the assembly of the most massive galaxies observed in the present-day Universe \citep[e.g.,][]{Dekel+2009a, Dekel+2009}. Studying quasar environments, particularly at $z\sim2-3$, near the peak of galaxy and BH formation \citep{Madau+2014, Richards+2006, Kulkarni+2019}, is therefore crucial to understanding the complex mechanisms regulating the baryon cycle in early massive galaxies. The {\it James Webb} Space Telescope (JWST) has been a game changer by enabling efficient and sensitive surveys of quasar fields down to $z\sim 7$, frequently revealing Mpc-scale structures of clustered galaxies \citep[see, e.g.,][]{Kashino+2023b, Eilers+2024}. However, the huge luminosity of quasars (up to $L_{\rm bol} > 10^{48}\,{\rm erg\,s^{-1}}$), can outshine any nearby sources within a few kiloparsecs, making detailed studies of their immediate surroundings challenging -- even for JWST \citep[e.g.][]{Marshall+2021, Ding+2022, Ding+2023, Stone+2023}. (Sub)millimeter observations are particularly effective in these cases, when the flux contrast between quasars and proximate sources is unfavorable. The Atacama Large (sub-)Millimeter Array (ALMA) has been pivotal in this regard, unveiling companion galaxies near quasars at cosmic noon \citep[see, e.g.][]{Bischetti+2018, Bischetti+2021, Fogasy+2020, Banerji+2021}, and beyond \citep[e.g,][]{Decarli+2017, Decarli+2019, Trakhtenbrot+2017, Neeleman+2019, Venemans+2019, Fudamoto+2021}. These studies often reveal tidal interactions and disturbed kinematics of the cold interstellar medium (ISM) -- either in the quasar host or its surroundings -- suggesting that such interactions may fuel BH growth and star formation at high redshift \citep[see, e.g.,][]{Alexander+2012}. Mergers and interactions can trigger both nuclear activity and starbursts \citep[see, e.g.][]{Urrutia+2008, Volonteri+2015, Angles-Alcazar+2017}, potentially driving powerful outflows that reshape galaxy properties \citep[see, e.g.][]{Carniani+2015b, Zakamska+2016, Bischetti+2017, Bischetti+2019a, Herrera-Camus+2020b}.

In this work, we report the discovery and characterization of a massive rotating disk galaxy near the hyperluminous quasar QSO CTS G18.01 at $z\approx3.25$ \citep[see][]{Deconto-Machado+2023}. The galaxy (hereafter MQN01-QC), lies $\sim 1''$ south to the quasar, within the MQN01 field \citep[MUSE Quasar Nebula 01; see][]{Borisova+2016}. This field hosts one of the highest known concentration of galaxies and AGN uncovered through deep multiwavelength surveys \citep[][]{Pensabene+2024, Galbiati+2025, Travascio+2025}, and found to be connected by Cosmic Web filaments revealed via Ly$\alpha$ emission across $\sim 4\,{\rm cMpc}$ (Cantalupo et al., in prep.). This system represents an ideal laboratory where to study the assembly of the progenitors of today's most massive galaxies in a dense region of the early Universe.

In \citet[][hereafter Paper I]{Pensabene+2024}, the first results were presented from an ALMA survey targeting tracers of galaxy ISM, including carbon monoxide (CO) rotational transitions and millimeter dust continuum, over $\sim 2\,{\rm arcmin^{2}}$ around the quasar. The relatively low resolution of these mosaic observations ($\sim1\rlap{.}{\arcsec}2$), caused the quasar host and its companion (previously dubbed as Object B in Paper I) to appear blended,  preventing spatially resolved analysis. Here, we present new results obtained from ALMA band 3 single-pointing observation targeting the CO(4--3) line at a much higher resolution ($\sim 0\rlap{.}{\arcsec}3$), toward the inner $\sim 150\,{\rm kpc}$ around the quasar.

This paper is structured as follow: in Sect.~\ref{sect:data-reduction} we describe the observations and the reduction of the data. In Sect.~\ref{sou-prop} we measure fluxes and luminosities and derive molecular gas masses and dust characteristics of the system. We present the morphological and kinematical analysis in Sect.~\ref{sect:kinematics}, and we discuss our results in Sect.~\ref{sect:discussion}. Finally, in Sect.~\ref{sect:conclusions} we summarize our findings and we draw conclusions.

Throughout this paper we assume a standard $\Lambda\rm{CDM}$ cosmology with $H_0=67.7\,\si{km\,s^{-1}Mpc^{-1}}$, $\Omega_{m}=0.310$, $\Omega_{\Lambda}=1-\Omega_{m}$ from \citet{PlanckColl+2020}. With these parameters and angular diameter of $1\arcsec$ corresponds to a physical scale of about $7.69\,{\rm kpc}$ at $z=3.25$.

\section{Observations and data processing}\label{sect:data-reduction}
\subsection{Description of ALMA observations}
In this work, we employed data from the ALMA Cycle 8 program 2021.1.00793.S (PI: S. Cantalupo) which includes ALMA band 3 and 6 low-resolution ($\sim 1\rlap{.}{\arcsec}2$) mosaics, as well as a band 3 single-pointing observation at higher angular resolution ($\sim 0\rlap{.}{\arcsec}3$) covering the central part of the MQN01 structure. The ALMA band 3 and 6 data were designed to target the CO(4--3) and CO(9--8) line (rest-frame frequencies $461.041$, and $1036.912\,{\rm GHz}$, respectively) in galaxies around $z\simeq3.2$, as well as the underlying 3- and 1.2-mm (rest-frame wavelengths $706$ and $282\,{\rm \mu m}$) dust continuum, respectively. Data acquisition and processing of the mosaics are described in Paper I to which we refer for full details. The single pointing observation is centered at the coordinates ICRS 00:41:32.273 -49:36:19.604. The Half Power Beam Width (HPBW) of individual ALMA 12-m antennas at the reference frequency of $109.00\,{\rm GHz}$ is $\approx 53\rlap{.}{\arcsec}4$. The frequency setup consists of four 1.875 GHz-wide spectral windows (SPWs). We disposed two adjacent SPWs in the upper side band (USB) centered respectively at $107.198\,{\rm GHz}$ and $109.00\,{\rm GHz}$, and other two  in the lower side band (LSB) covering the frequency range $94.06-97.74\,{\rm GHz}$ contiguously. The native spectral resolution of the acquired data is $1.95\,{\rm MHz}\,{\rm channel^{-1}}$ ($\sim 5.4\,{\rm km\,s^{-1}}\,{\rm channel^{-1}}$ at $109.00\,{\rm GHz}$). The observations were executed in four blocks during the period 2021 November 13--14, employing a total on-source exposure time of $\approx 4.97\,{\rm h}$. The number of ALMA main array antennas ranged from $42$ to $49$, with baselines within $41.4-3638.2\,{\rm m}$. The weather conditions during executions provided a mean precipitable water vapor in the range $0.7-1.2\,{\rm mm}$. The nominal achieved sensitivity is $64\,{\rm \mu Jy\,beam^{-1}}$ over a bandwidth of $36.358\,{\rm MHz}$ ($\simeq 100\,{\rm km\,s^{-1}}$), and $4.5\,{\rm \mu Jy\,beam^{-1}}$ over the aggregate continuum in the $6.78\,{\rm GHz}$ bandwidth. During the observations, the quasars J0025-4803, J2357-5311 were employed as a phase and flux calibrator, respectively. 

  \begin{figure*}[!ht]
   	\centering
   	\resizebox{\hsize}{!}{
		\includegraphics{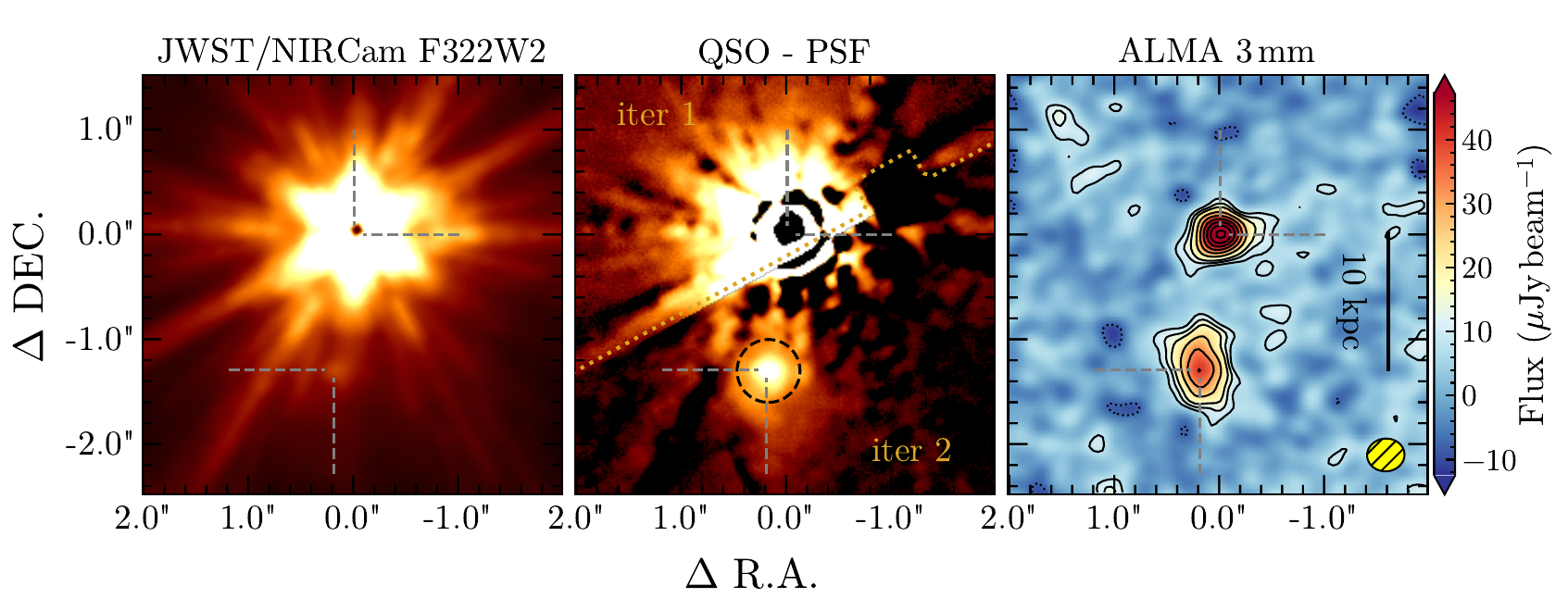}}
       \caption{Images of the MQN01-QC galaxy located $\sim 10\,{\rm kpc}$ south to the hyperluminous quasar QSO CTS G18.01 at $z\approx3.2$. {\it Left panel:} {JWST}/NIRCam $F322W2$ snapshot \citep{Wang+2025}. {\it Central panel:} same as the left panel but with two rounds of quasar light removal (QSO - PSF). The result, shown in the lower right half of the image below the dotted demarcation line, provides a remarkably clearer view of the galaxy (indicated by the circle) compared to the original image on the left. Details about the light removal and stellar mass estimation are described in Sect.~\ref{ssect:qso-psf} and ~\ref{ssect:mstar}, respectively. Similar steps were conducted to the NIRCam $F150W2$ filter image.  {\it Right panel:} ALMA 3-mm dust continuum image. The gray dashed lines indicate the nominal location of the sources as derived from the peak of the dust continuum. The dotted and solid contours in the ALMA map correspond to $-2\sigma$ and $[3,2n]\sigma$, respectively, where $n>1$ is an integer, and $\sigma$  denotes the noise RMS. The ALMA synthesized beam is shown as a yellow ellipse in the bottom right corner.}
       \label{fig:det-maps}
    \end{figure*}
    
\subsection{Reduction and imaging of ALMA data}
\label{ssect:data-reduction}
We performed the reduction and processing of the raw visibilities by utilizing the Common Astronomy Software Application \citep[CASA;][]{McMullin+2007,Hunter+2023}. We performed standard calibration of the measurement sets by running the \texttt{scriptForPI} delivered by the ALMA observatory, using the CASA pipeline version $6.2.1$. We then imaged the ALMA band 3 visibilities with the CASA task \texttt{tclean} employing the H\"{o}gbom deconvolution algorithm \citep{Hogbom1974}, and adopting a ``natural'' weighting scheme of the visibilities to maximize the surface brightness sensitivity per beam. This procedure yielded a naturally weighted beam size and position angle at $109.00\,{\rm GHz}$ of $0\rlap{.}{\arcsec}32\times0\rlap{.}{\arcsec}29$ and $-86.18\,{\rm deg}$, respectively. The resulted angular resolution corresponds to $\simeq 2.3\,{\rm kpc}$ at $z=3.25$. 

We made an initial inspection of the data obtaining a ``dirty'' datacube by Fourier-transforming the raw visibilities with the \texttt{tclean} task setting \texttt{niter=0}, \texttt{specmode=``cube''} as a spectral definition mode, and a channel width of $25\,{\rm km\,s^{-1}}$. Similarly, we obtained ``dirty'' 3-mm continuum image by aggregating all the four SPWs by running \texttt{tclean} with \texttt{niter=0} in multifrequency synthesis mode (\texttt{specmode=``mfs''}). We imaged the visibilities by using a pixel size of $0\rlap{.}{\arcsec}04$, thus achieving a sufficient sampling the beam minor axis with $\sim 7$ pixels. 

In this dataset, we detected the quasar host and the MQN01-QC companion galaxy, in both their CO(4--3) line and the underlying 3-mm dust continuum. They are revealed with an on-sky projected separation of $\sim 1''$ (see Fig.~\ref{fig:det-maps}). No additional galaxy emission was detected in the field-of-view (FoV). Hence, we obtained the ``cleaned'' datacube and line-free continuum image by cleaning down to a threshold of  $1.5\sigma$ within two circular masks of $0\rlap{.}{\arcsec}6$ radius. These were sufficient to encompass the observable emission of both the sources. Finally, we obtained a continuum-subtracted datacube by employing the CASA task \texttt{imcontsub}. In this procedure, we fit with a zeroth-order polynomial and subtract the line-free channels in the image domain within $20\arcsec\times20\arcsec$ rectangular mask around the sources.

\subsection{Ancillary datasets}\label{ssect:other-data}
In this work, we capitalize on the low-resolution ALMA band 6 mosaic observations toward MQN01 field reported in Paper I. This includes cleaned continuum image and datacube with a channel width of $40\,{\rm km\,s^{-1}}$, and naturally weighted beam sizes of $1\rlap{.}{\arcsec}3\times1\rlap{.}{\arcsec}1$, and $1\rlap{.}{\arcsec}2\times1\rlap{.}{\arcsec}0$ at the reference frequency of $245.00\,\rm{GHz}$, respectively. In this dataset, we detected the continuum emission at 1.2 mm from both the sources, and the CO(9--8) line from the quasar host galaxy.

Additionally, we benefit from photometric observations acquired by the JWST/NIRCam \citep{Rigby+2023}, as a part of the program GO 1835 \citep[PI: Cantalupo;][]{Wang+2025}, using the extra-wide filters $F150W2$ and $F322W2$ in the short- and long-wavelength channel, respectively. These images cover a FoV of $2\times 5\,{\rm arcmin^{2}}$, encompassing the sky area of the ALMA observations, with an on-source exposure time of 1632 sec per filter.  

\subsection{Quasar light removal from JWST/NIRCam images}\label{ssect:qso-psf}
In the acquired JWST images, the MQN01-QC galaxy appears outshone by the central quasar emission but still visible in both NIRCam filters (see Fig.~\ref{fig:det-maps}). We performed optimal subtraction of the quasar point-spread function (PSF) by employing a bright star in the field. The alignment and flux scaling of the star were made by comparing the locations and brightnesses of the light spikes of the star and quasar in each filter \citep{Ren+2024}.  A residual light was present in the cutout around the quasar after this initial round of PSF removal (``iter 1'' in Fig.~\ref{fig:det-maps}, central panel). These residuals are most likely caused by the mismatch between the quasar and the star light profile due to their different spectral shapes \citep{Ren+2024}. In order to minimize the light residuals, we mirrored the upper half of the image where no bright additional source is apparent, and subtracted it from the lower half where the galaxy is located. {\rm We note that this step also removes the large-scale background of any nature.} We report the resulting image as "iter 2" in Fig.~\ref{fig:det-maps} (central panel), which provides remarkably clearer view of the galaxy and its extended morphology compared to the original image (left panel). We conducted similar steps on the NIRCam $F150W2$ filter image.

\begin{table*}[!htbp]
\def\arraystretch{1.15}
\caption{Line and continuum flux measurements, luminosity estimates, and derived quantities of the sources.}  
\label{tbl:sou_flux_prop}      
\centering 
\resizebox{0.9\hsize}{!}{
\begin{tabular}{l | ccc | cc}     
\toprule\toprule
										&\multicolumn{3}{c|}{QSO CTS G18.01}			&\multicolumn{2}{c}{MQN01-QC} \\
\cmidrule(lr){1-6}
R.A. (ICRS) 								&\multicolumn{3}{c|}{00:41:31.443} 				&\multicolumn{2}{c}{00:41:31.463} 	\\
Dec. (ICRS) 								&\multicolumn{3}{c|}{-49:36:11.703} 				&\multicolumn{2}{c}{-49:36:12.943} 	\\
\bottomrule
Line$\,^{(1)}$										&\multicolumn{2}{c}{CO(4--3)} 							&{CO(9--8)}					&CO(4--3)$\,^{(\dagger)}$ 						&CO(9--8)$\,^{(\ddagger)}$		\\
	& systemic comp. & blueshifted broad comp. &	&	&	\\
\cmidrule(lr){1-6}
$\Delta\varv_{\rm QSO}\,({\rm km\,s^{-1}})$ 		&$-$								&$-307^{+125}_{-137}$				&$-36^{+40}_{-44}$				&$-241^{+77}_{-70}$					&$-$					\\
$z_{\rm CO}$								&$3.2510^{+0.0004}_{-0.0004}$		&$3.2468^{+0.0018}_{-0.0022}$		&$3.2504^{+0.0005}_{-0.0004}$	&$3.2475^{+0.0012}_{-0.0010}$			&$-$					\\
FWHM$_{\rm CO}\,({\rm km\,s^{-1}})$			&$404^{+64}_{-66}$					&$717^{+225}_{-187}$				&$556^{+128}_{-115}$			&$796^{+55}_{-44}$					&$-$					\\
$F_{\rm CO}\,({\rm Jy\,km\,s^{-1}})$$\,^{(1)}$		&$0.48^{+0.13}_{-0.16}$  				&$0.32^{+0.17}_{-0.14}$ 				&$1.06^{+0.20}_{-0.18}$ 			&$1.1^{+0.3}_{-0.3}$					&$<0.3$				\\
$L_{\rm CO}\,(10^{8}\,L_{\odot})$ 				&$0.44^{+0.12}_{-0.15}$				&$0.30^{+0.16}_{-0.13}$				&$2.2^{+0.4}_{-0.4}$				&$1.0^{+0.3}_{-0.2}$					&$<0.3$				\\
$L'_{\rm CO}\,(10^{10}\,{\rm K\,km\,s^{-1}\,pc^{2}})$  &$1.4^{+0.4}_{-0.5}$					&$1.0^{+0.5}_{-0.4}$					&$0.62^{+0.12}_{-0.10}$			&$3.3^{+0.8}_{-0.8}$					&$<0.9$				\\
\bottomrule
Continuum$\,^{(2)}$							&\multicolumn{2}{c}{${\rm 3\,mm}$}		&${\rm1.2\,mm}$	&${\rm 3\,mm}$				&${\rm1.2\,mm}$		\\
\cmidrule(lr){1-6} 	 					
$F_{\rm peak}\,({\rm mJy\,beam^{-1}})$			&\multicolumn{2}{c}{$0.072\pm0.004$}	&$0.99\pm0.04$	&$0.043\pm0.004$			&$1.63\pm0.04$		\\
$F_{2\sigma}\,({\rm mJy})$					&\multicolumn{2}{c}{$0.092\pm0.008$}	&$-$				&$0.091\pm0.008$			&$-$					\\
\bottomrule
\end{tabular}
}
\tablefoot{$\,^{(1)}$In the case of the CO(4--3) line of the QSO host and MQN01-QC galaxy, best-fit parameters of the Gaussian components in the composite model are reported. $\,^{(2)}$The measurements of the 1.2-mm continuum fluxes are taken from Paper I, where the low spatial resolution of the band 6 mosaic observations makes the emission from the two sources partially blended. The reported measurements are obtained through a single-pixel analysis, thus minimizing the flux contamination. $\,^{(\dagger)}${\rm The parameter estimates reported in this column are the flux-weighted average velocity shift and redshift, the total full-width at half maximum, and the flux and luminosity of the best-fit composite model of the line profile, respectively.} $^{(\ddagger)}$The reported values correspond to $3\sigma$ upper limits assuming $FWHM=750\,{\rm km\,s^{-1}}$.}
\end{table*}

    \begin{figure}[!ht]
   	\centering
   	\resizebox{\hsize}{!}{
		\includegraphics{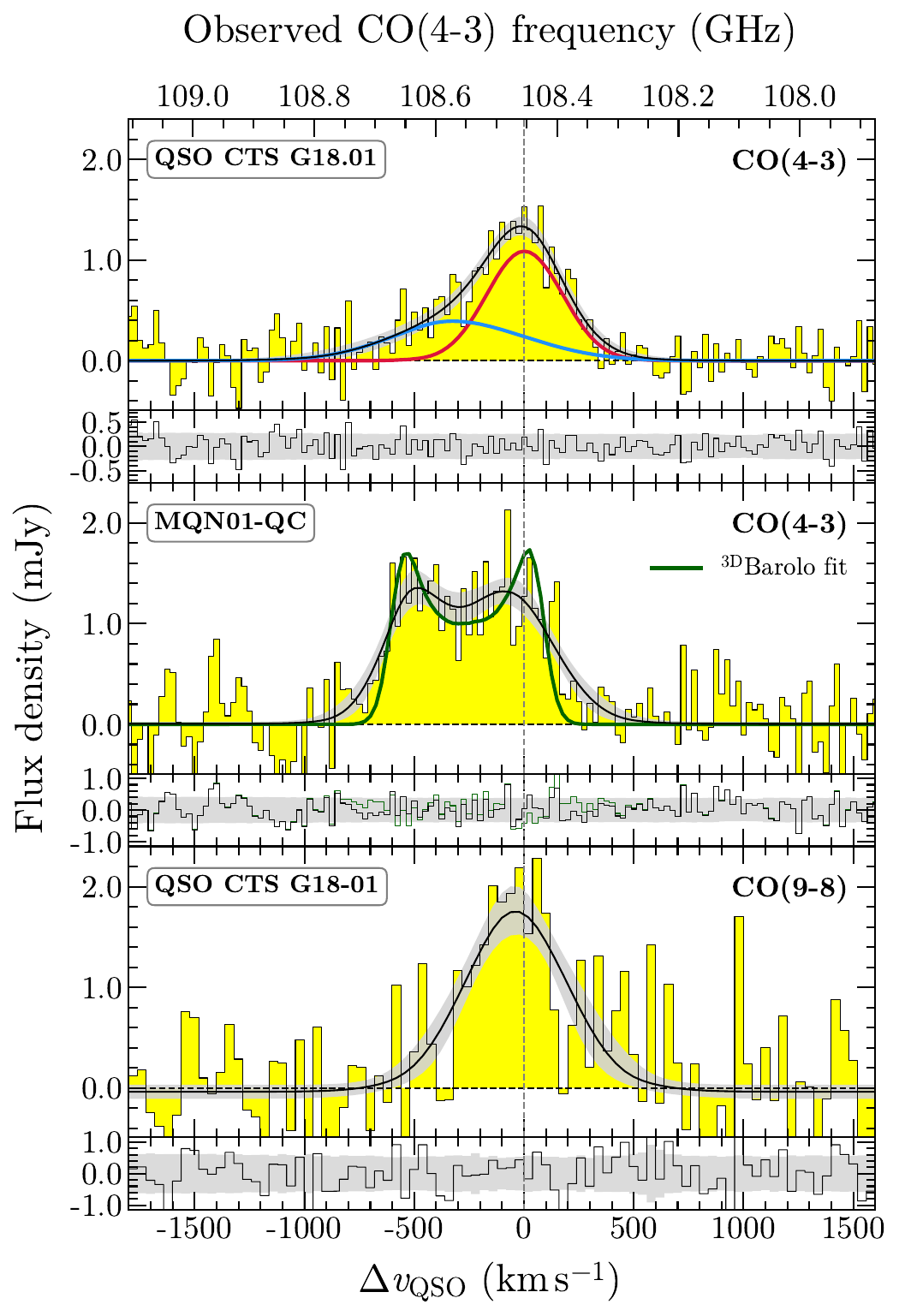}}
       \caption{Continuum-subtracted CO(4--3) and CO(9--8) line of the QSO CTS G18.01 host galaxy ({\it top and bottom panel}), and the CO(4--3) line of the MQN01-QC ({\it central panel}). The data are reported in yellow. The black solid line shows the best-fit composite model. The gray shaded area shows the $1\sigma$ confidence interval of the best fit model. The individual Gaussian components are reported in red and blue for the quasar line profile. The green solid line in the central panel shows the line profile extracted from the best-fit rotating disk model as obtained with $^{\rm 3D}${\sc Barolo} (see Sect.~\ref{sect:kinematics}). The fit residuals (data $-$ model) are also reported below each spectra, where the shaded area represents the noise RMS. The bottom velocity scale refers to the quasar systemic redshift as determined by the red component of the quasar CO(4--3) line profile.}
       \label{fig:spectra}
    \end{figure}

     \begin{figure*}[!ht]
   	\centering
   	\resizebox{\hsize}{!}{
		\includegraphics{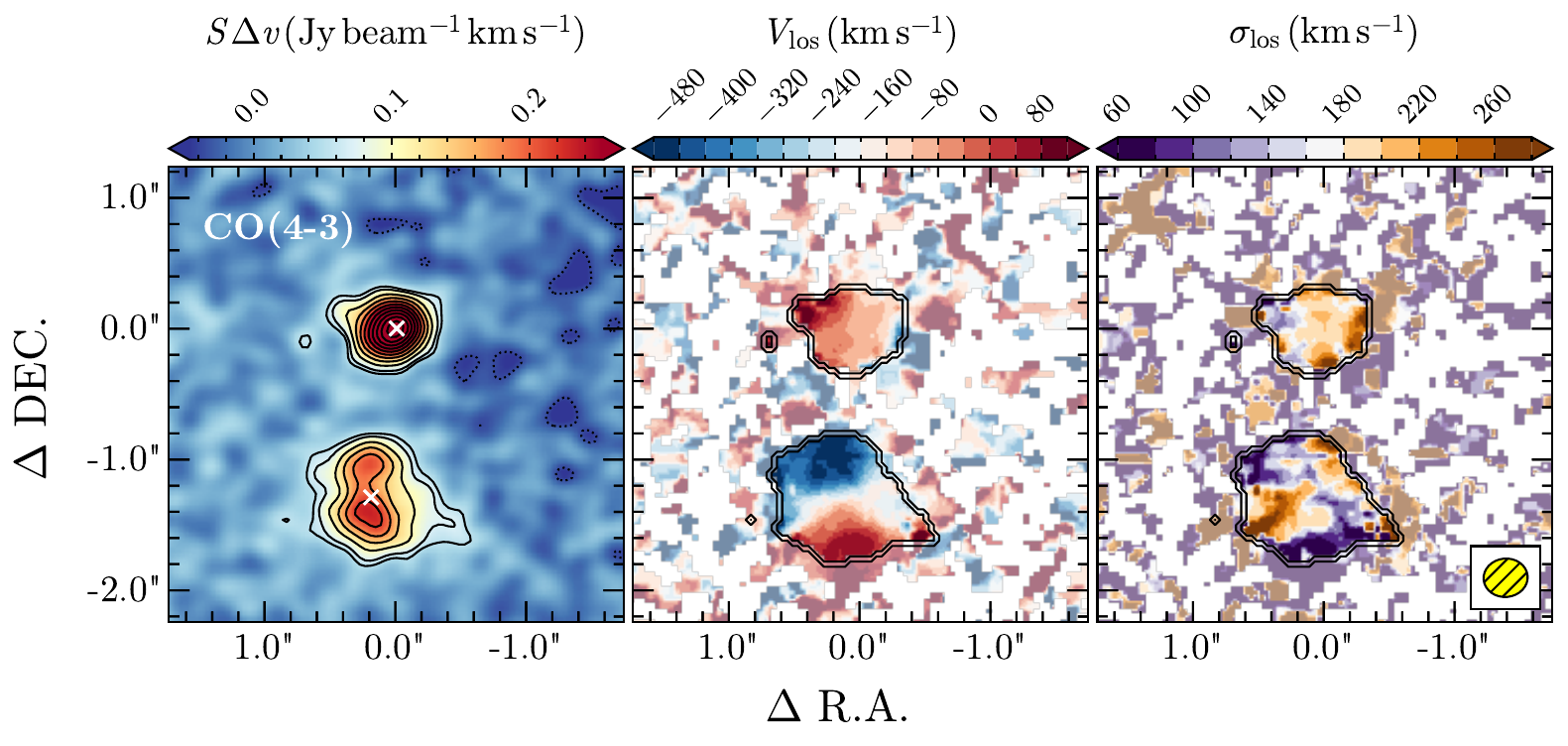}}
       \caption{The CO(4--3) line-velocity integrated map (moment 0th, {\it left panel}), the line-of-sight velocity field ($V_{\rm los}$, moment 1st, {\it central panel}), and the line-of-sight velocity dispersion field ($\sigma_{\rm los}$, moment 2nd, {\it right panel}) of the quasar host--MQN01-QC galaxy system. These maps are obtained from the continuum-subtracted ALMA datacube using channels within $[-700;+400]\,{\rm km\,s^{-1}}$ around the quasar systemic redshift ($V_{\rm los}=0$). The dotted (negative) and solid (positive) contours correspond to $[-2,3,2n]\sigma$, where $n\ge2$ is an integer, and $\sigma$ denotes the noise RMS. The white crosses mark  the peak emission of the 3-mm dust continuum. The $3\sigma$ contour of the moment 0th is also reported in the moment 1st and 2nd maps. The ALMA synthesized beam is shown in the bottom right corner.}
       \label{fig:mom-maps}
    \end{figure*}
    
\section{Characterization of the quasar host--companion galaxy system}\label{sou-prop} 
\subsection{Flux and luminosity measurements from ALMA data}\label{sect:sou-flux}
From the ALMA band 3 cleaned datacube we extracted the CO(4--3) line emission of the sources by applying the $2\sigma$-clipped photometry method \citep[see, e.g.,][]{Bethermin+2020, Pensabene+2024} which enables to recover line flux with optimal signal-to-noise ratio (${\rm S/N}$) while preventing significant flux losses. We first extracted the spectrum of the sources from the continuum-subtracted cleaned datacube, at the peak pixel of the 3-mm dust continuum emission, corrected for the response of the primary beam (PB), and fitted the line using a single Gaussian component. Then, we obtained the line-velocity integrated map (moment 0th) by summing all the channels within $\pm2\sigma$ from centroid of the extracted line profile. Hence, we extracted a new spectrum including the signal from all the pixels encompassed by the ${\rm S/N \ge 2}$ isophote, and multiplying for the square-root of the independent beam elements in the extracting region. Finally, we repeated the line fit on this new spectrum as described above, and we iterated the procedure until convergence. 

The observed CO(4--3) line profiles of our sources cannot be adequately reproduced with a single Gaussian component. Indeed, the CO(4--3) line of the quasar host galaxy exhibits a broad blueshifted tail that need to be modeled (see Fig.~\ref{fig:spectra}). This feature suggests the presence of a non-circular gas kinematic component. On the other hand, the CO(4--3) line of MQN01-QC exhibits a boxy double-peaked profile. This type of line profile is often observed in rotating disks with significant inclination angle \citep[see, e.g.,][]{Elitzur+2012, deBlok+2014, Kohandel+2019, Kohandel+2024}. 

We inspected the ALMA band 6 datacube and detect spatially unresolved CO(9--8) line emission from the quasar host galaxy. In this case, we measured the line flux by applying a standard single-pixel analysis. We therefore extracted the line flux from the peak pixel of the emission and corrected for the PB response at the corresponding location. We report the extracted spectra in Fig.~\ref{fig:spectra}. Interestingly, the CO(9--8) line of the quasar host lacks of similar blushifted wing as observed in the CO(4--3) line profile\footnote{Assuming the same flux ratio between the systemic and broad component of the CO(4--3) line (see Table~\ref{tbl:sou_flux_prop}), a broad blushifted CO(9--8) component with ${\rm FWHM=700\,{\rm km\,s^{-1}}}$ is expected to have a peak flux of $\simeq 0.9\,{\rm mJy}$ and therefore be detectable at the current sensitivity in the ALMA band 6 data in the spectrum of the quasar host galaxy.}. This indicates that the latter component is associated to molecular gas exposed to different physical conditions and excitation mechanisms than that traced by the bulk of the CO(4--3) emission -- which is expected to be retained within the ISM of the quasar host galaxy. Indeed, while the low-$J$ CO(4--3) transition is a tracer of the cold molecular gas, the CO(9--8) is a high-$J$ CO rotational transition which is typically associated to high-density\footnote{The critical density, that is the rate at which collisional de-excitations equals the spontaneous radiative decay rate of the CO(4--3) and CO(9--8) transitions is $n_{\rm crit}\simeq8.7\times10^{4}\,{\rm cm^{-3}}$ and $\simeq 8.7\times10^{5}\,{\rm cm^{-3}}$, respectively \citep{Schoier+2005}.} and/or high-temperature molecular gas exposed to strong X-ray radiation field or heated by shocks produced by AGN-driven outflows \citep[e.g.,][]{Maloney+1996, Meijerink+2013, Vallini+2019, Pensabene+2021}. 

Following the aforementioned considerations, we performed a more accurate fit of the CO line profiles by employing a composite double and single Gaussian model for the CO(4--3) and CO(9--8) lines, respectively, using the Python Markov chain Monte Carlo (MCMC) ensemble sampler \texttt{emcee} package \citep{Foreman+2013}. In the fitting procedure, we adopted Gaussian uncertainties in the definition of the likelihood and flat priors on the free parameters such as amplitudes, FWHM, and line centroids. Finally, we derived the line luminosities as \citep[see, e.g.,][]{Solomon+1997}
\begin{align}
L_{\rm CO}\,\quadra{L_{\astrosun}}&=1.04\times10^{-3}\,F_{\rm CO}\,\nu_{\rm obs}\,D_{L}^{2},\\
L'_{\rm CO}\,\quadra{\rm K\,km\,s^{-1}pc^{2}}&=3.25\times10^{7}\,F_{\rm CO}\,\frac{D_{L}}{(1+z)^{3}\,\nu_{\rm obs}^{2}},
\end{align}
where $F_{\rm CO}$ is the velocity-integrated line flux {\rm obtained with the Gaussian fit (see Table~\ref{tbl:sou_flux_prop})} in units of ${\rm Jy\, km\,s^{-1}}$, $\nu_{\rm obs}=\nu_{\rm rest}/(1+z)$ is the observed central frequency of the line in GHz at the redshift $z$, and $D_{L}(z)$ is the luminosity distance in Mpc. The two luminosity measurements are related via $L_{\rm CO}=3\times10^{-11}\,\nu_{\rm rest}^{3}\,L'_{\rm CO}$. We report best-fit models, parameters, and the derived quantities in Fig.~\ref{fig:spectra}, and Table~\ref{tbl:sou_flux_prop}. For the subsequent analysis, we adopt as systemic velocity of the system, the observed line centroid of the narrow component of the quasar CO(4--3) line profile.

We measured the 3-mm continuum flux of the sources. To do so, we added together the fluxes of individual pixels above the ${\rm S/N \ge 2}$ isophote in the continuum map, divided by the beam size in pixels, and corrected the total flux for the PB response at the location of the galaxies. We also computed the flux uncertainties by rescaling the noise root-mean-square (RMS) of the image by the PB response, and multiplying by the square-root of the number of independent beams within the ${\rm S/N \ge 2}$ region. For sake of completeness, in Table~\ref{tbl:sou_flux_prop} we report both the 3-mm and 1.2-mm continuum measurements derived in this work and in Paper I, respectively. 

\subsection{CO(4--3) line flux distributions and velocity fields}
In Fig.~\ref{fig:mom-maps}, we show the moment maps of the system. Specifically, we obtained the line velocity-integrate map (moment 0th) by integrating channels within $[-700;+400]\,{\rm km\,s^{-1}}$ around the quasar systemic redshift, in order to include all the CO(4--3) emission from both the quasar host and MQN01-QC galaxy (see Fig.~\ref{fig:spectra}). We obtained the line-of-sight velocity field (moment 1st), and the velocity dispersion field (moment 2nd), by collapsing channels after masking voxels in the cube with ${\rm S/N < 2}$. 

As it is evident from Fig.~\ref{fig:mom-maps}, our analysis reveals very distinct characteristics of the two sources. The CO(4--3) line emission of the quasar host is relatively compact in the observed size and shows a radially declining flux distribution, similar to the morphology observed in the dust continuum at 3 mm (see Fig.~\ref{fig:det-maps}). The velocity field of the quasar host appears irregular. Although there is a hint of a velocity gradient along the E-W direction, with the current data it is difficult to determine with certainty if the observed gradient is a signature of rotation. The emission line of the MQN01-QC galaxy is instead resolved into two separated ``clumps'', deviating from the morphology revealed in the continuum. This suggests that the molecular gas throughout the galaxy disk is distributed in a complex structures -- such as, e.g., spiral arms. Similar structures have been also highlighted by other (sub-)millimeter observations of high-redshift galaxies and quasar hosts \citep[see, e.g.,][]{Walter+2004, Riechers+2009, Tacconi+2010, Hodge+2012, Hodge+2019, Iono+2016, Tadaki+2018, Roman-Oliveira+2023}.  Alternatively, the observed morphology as traced by the CO(4--3) emission of the MQN01-QC galaxy could result from significant molecular gas depletion or optically thick line emission in the galaxy central region compared to its outskirts. Furthermore, Fig.~\ref{fig:mom-maps} reveals a possible elongation of the cold gas distribution on the west side of the MQN01-QC disk, which may suggest that part of the gas is being stripped, possibly due to the interaction with the surrounding environment. However, this feature seems to do not significantly impact the global ordered kinematics characterizing the velocity field. Notably, the gas kinematics of the MQN01-QC galaxy shows a clear velocity gradients along the south-north direction which, at a first guess, is totally consistent with rotation. Finally, the observed velocity dispersion, although not yet corrected for beam smearing effect, is relatively low (see Sect.~\ref{sssect:kin-model}). All these lines of evidence indicates that the molecular gas in the MQN01-QC galaxy is settled in a rotationally-supported disk. This is somehow surprising given the small projected separation and, therefore, the possible interaction with the quasar host galaxy in its vicinity (see Sects.~\ref{ssect:dynamics}).

\subsection{Molecular gas mass and dust characteristics}\label{sect:md-mh2}
From the available CO(4--3) line luminosities we estimated the molecular gas content of the galaxies. To this purpose, assumptions on the CO(4--3)-to-CO(1--0) conversion factor ($L'_{\rm CO(1-0)} = L'_{\rm CO(4-3)}/r_{41}$), and the $\alpha_{\rm CO}$ coefficient are needed ($M_{\rm H2}=\alpha_{\rm CO}L'_{\rm CO(1-0)}$; see \citealt{Bolatto+2013}, for a review). {\rm Here, we adopt a recent estimate of the CO-to-H$_{2}$ conversion factor derived for a sample of local ultra-luminous infrared galaxies as $\alpha_{\rm CO} = 1.7\pm0.5\,M_{\astrosun}\,({\rm K\,km\,s^{-1}\,pc^{2}})^{-1}$ \citep{MontoyaArroyave+2023}, and $r_{\rm 41} = 0.87$ typical of quasar hosts and SMGs \citep{CarilliWalter2013}, and obtain $M_{\rm H2}^{\rm QC}=6^{+3}_{-2} \times 10^{10}\,(\alpha_{\rm CO}/1.7)\,M_{\astrosun}$, and $M_{\rm H2}^{\rm QSO}= 4.6^{+2.0}_{-1.8} \times 10^{10}\,(\alpha_{\rm CO}/1.7)\,M_{\astrosun}$ for the MQN01-QC and the quasar host galaxy, respectively. The latter decreases to $M_{\rm H2}^{\rm QSO, sys}= 2.5^{+1.7}_{-1.4} \times 10^{10}\,(\alpha_{\rm CO}/1.7)\,M_{\astrosun}$ when accounting only for the systemic component of the quasar CO(4--3) line profile. Finally, we provide an estimate of the bulk of the molecular gas mass in the quasar host galaxy. To do so, we employed the total CO line luminosity and subtracted the fraction associated to the blueside ($<-300{\rm km\,s^{-1}}$) of the broad line component. Thus, by adopting the same assumptions as above we obtained $M_{\rm H2}^{\rm QSO, bulk}= 3.5^{+2.0}_{-1.4} \times 10^{10}\,(\alpha_{\rm CO}/1.7)\,M_{\astrosun}$. We report these estimates in Table~\ref{tbl:budgets}. Here, we stress that the reported uncertainties on molecular gas masses do not take into account systematic uncertainties on the assumed conversion factors.}

We estimated the dust mass ($M_{\rm dust}$) of galaxies by performing the modeling of the Rayleigh-Jeans (RJ) tail of dust spectral energy distribution (SED) traced by the available 3- and 1.2-mm continuum measurements. To this purpose, we assumed a modified black body emission of the dust grains in optically thin limit \citep[see, e.g.,][]{Beelen+2006, Casey+2012, Leipski+2013, Decarli+2022}, and took into account the contrast effect due to the cosmic microwave background via the relation \citep[see][]{daCunha+2013, Zhang+2016}:
\eq{F_{\nu/(1+z)}^{\rm obs} = \frac{1+z}{D_{L}^{2}}\,M_{\rm dust}\,\kappa_{\nu}\tonda{B_{\nu}[T_{\rm dust}(z)]-B_{\nu}[T_{\rm CMB}(z)]}.\label{eq:bb}}
In Eq.~(\ref{eq:bb}), $T_{\rm dust}(z)$, and $T_{\rm CMB}(z)$, are the temperature of the dust and the CMB at redshift $z$, respectively, and $\kappa_{\nu}$ is the dust mass opacity coefficient defined as $\kappa_{\nu} = \kappa_{0}(\nu/\nu_{0})^{\beta}$ where $\beta$ is the dust spectral index. Here, we assumed $\kappa_\nu = 0.077\,{\rm m^{2}\,kg^{-1}}$ at $850\,{\rm \mu m}$ from \citet{Dunne+2000}.

\begin{figure}[!t]
   	\centering
   	\resizebox{\hsize}{!}{
		\includegraphics{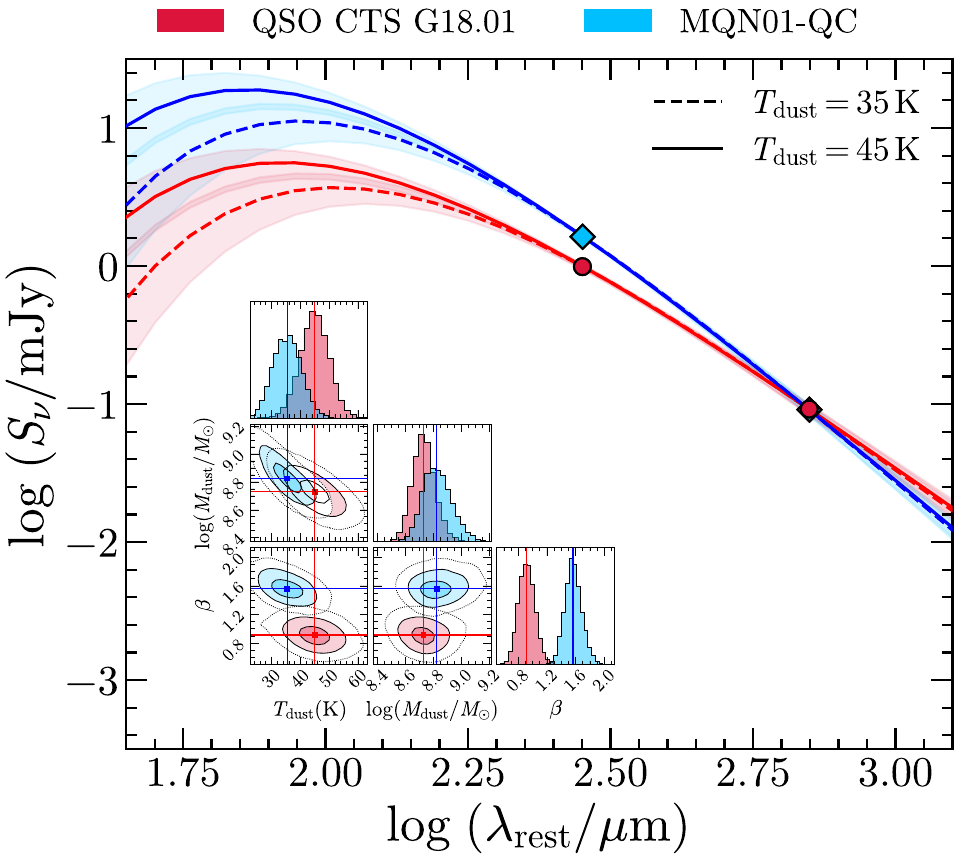}}
       \caption{Dust SED modeling of the quasar host (red circles) and the MQN01-QC galaxy (blue diamonds). The solid and dashed lines show the best-fit modified blackbody models for different assumptions on the dust temperature. The $1\sigma$ confidence intervals on the best fit are reported with shaded areas. The inset corner plot shows the posterior probability distributions of the free parameters for the quasar host (red) for $T_{\rm dust}=45\,{\rm K}$, and the MQN01-QC (blue) for $T_{\rm dust}=35\,{\rm K}$.}
       \label{fig:dust-seds}
    \end{figure}

We performed the fit of the dust RJ tail by minimizing the log-likelihood function assuming Gaussian uncertainties on measurements, and exploring the parameter space through the {\tt emcee} \citep{Foreman+2013} Python package, adopting flat priors on $M_{\rm dust}$ and $\beta$. Due to the lack of continuum measurements near to the peak of the dust SED, we are able to put only shallow constraints on the dust temperature. For this reason, we performed the fit by assuming two different Gaussian priors for the $T_{\rm dust}$, the value of which we centered at $35\,{\rm K}$, typical of $z\sim1-3.5$ sub-millimeter galaxies \citep[see, e.g.,][]{Chapman+2005, Kovacs+2006, Dudzeviciute+2020}, and at $45\,{\rm K}$, typical of $z\sim2-5$ quasar host galaxies \citep[see, e.g.,][]{Duras+2017, Bischetti+2021, Tripodi+2024}, assuming a standard deviation on the dust temperature priors as $\sigma(T_{\rm dust})=5\,{\rm K}$. In Fig~\ref{fig:dust-seds}, we show the data, the best-fit models as well as the posterior distributions of free parameters. We report our results in Table~\ref{tbl:budgets}. 

We computed the total infrared luminosities ($L_{\rm FIR}$) by integrating the best-fit dust SED models in the rest-frame wavelength range $8-1000\,{\rm \mu m}$, and we inferred the IR-based star-formation rate by using the scaling relation for local galaxies \citep{Murphy+2011}\footnote{This recipe holds under the hypothesis that the entire Balmer continuum (i.e., $912\,\AA < \lambda < 3646\,\AA$) is absorbed and re-irradiated by the dust in optically-thin limit. Additionally, a Kroupa initial mass function \citep{Kroupa2001} is implicitly assumed, having a slope of $-1.3$, and $-2.3$ for stellar masses between $0.1-0.5\,M_{\astrosun}$, and $0.5-100\,M_{\astrosun}$.}: ${\rm SFR}_{\rm IR}\,(M_{\astrosun}\,{\rm yr^{-1}})=1.49\times10^{-10}L_{\rm FIR}/L_{\astrosun}$. We note that, under our working assumptions, any possible contribution from the central AGN to the measured FIR luminosity would result in an overestimation of the ${\rm SFR}_{\rm IR}$. For the subsequent analysis,, we adopt fiducial values assuming $T_{\rm dust} = 45\,{\rm K}$ for the quasar host, and $T_{\rm dust} = 35\,{\rm K}$ for MQN01-QC, in line with expectation from previous aforementioned works.

Finally, we modeled the 3-mm continuum morphology of both galaxies using a 2D elliptical Gaussian by running the CASA task {\tt imfit}. This simple model is sufficient to reproduce well the observed dust distribution in both galaxies. As a result, we found an intrinsic (beam-deconvolved) 3-mm half-light radius\footnote{The half-light radius for a Gaussian distribution is equivalent to half of the FWHM of the major axis.} of $R_{\rm 1/2,\,3mm}=0.14\pm0.03\,{\rm arcsec}$ (or $1.1\pm0.2\,{\rm kpc}$), and $0.29\pm0.04\,{\rm arcsec}$ (or $2.2\pm0.3\,{\rm kpc}$) for the quasar host and MQN01-QC galaxy, respectively. We report these measurements in Table~\ref{tbl:budgets}.

\begin{table}[!t]
\def\arraystretch{1.15}
\caption{Mass budgets, dust characteristics, and morphological sizes of the quasar host and the MQN01-QC galaxy.}  
\label{tbl:budgets}      
\centering 
\resizebox{\hsize}{!}{
\begin{tabular}{lcc}       
\toprule\toprule
	&					QSO CTS G18.01			&		MQN01-QC\\
\cmidrule(lr){1-3}
\multirow{2}{*}{$M_{\rm dyn}(<r)$ ($10^{11}\,M_{\astrosun}$)}			&	{\rm disp:}$\,0.13^{+0.06}_{-0.04}$ &	\multirow{2}{*}{$2.5^{+0.6}_{-1.1}\,(<4.1\,{\rm kpc})$} \\ &{\rm rot:}$\,0.7^{+0.8}_{-0.3}\, (<1.3\,{\rm kpc})$$\,^{(\dagger)}$ & \\
\cmidrule(lr){1-3}	
\multirow{1}{*}{$M_{\star}(<r)$ ($10^{10}\,M_{\astrosun}$)}		&	$-$										&	$9.1^{+1.4}_{-1.1}\,(<2.3\,{\rm kpc})$\\
\cmidrule(lr){1-3}
\multirow{3}{*}{$M_{\rm H2}$$\,^{(1)}$ ($10^{10}\,(\alpha_{\rm CO}/1.7)\,M_{\astrosun}$)}		&{\rm total:}$\,4.6^{+2.0}_{-1.8}$\\  &{\rm sys:}$\,2.5^{+1.7}_{-1.4}$ &\multirow{1}{*}{$6^{+3}_{-2}$}\\  &{\rm bulk:}$\,3.5^{+2.0}_{-1.4}$ 		\\
\cmidrule(lr){1-3}
$T_{\rm dust}$$\,^{(2)}$(K)						&\multicolumn{2}{c}{$\mathbf{35\pm5}$}			\\	
$\log(M_{\rm dust}/M_{\astrosun})$					&$8.88^{+0.11}_{-0.09}$		&$\mathbf{8.83^{+0.11}_{-0.09}}$\\
$\beta$										&$1.01^{+0.13}_{-0.12}$		&$\mathbf{1.56^{+0.12}_{-0.11}}$\\
$L_{\rm FIR}$$\,^{(3)}$ ($10^{13}\,L_{\astrosun}$)		&$0.08^{+0.04}_{-0.03}$		&$\mathbf{0.27^{+0.15}_{-0.10}}$\\
SFR$_{\rm IR}$ $(M_{\astrosun}\,{\rm yr^{-1}})$		&$124^{+60}_{-41}$			&$\mathbf{405^{+216}_{-149}}$\\
\cmidrule(lr){1-3}
$T_{\rm dust}$$\,^{(2)}$(K)						&\multicolumn{2}{c}{$\mathbf{45\pm5}$}					\\
$\log(M_{\rm dust}/M_{\astrosun})$					&$\mathbf{8.73^{+0.08}_{-0.08}}$		&$8.68^{+0.08}_{-0.08}$\\
$\beta$										&$\mathbf{0.92^{+0.12}_{-0.11}}$		&$1.47^{+0.11}_{-0.11}$\\
$L_{\rm FIR}$$\,^{(3)}$ ($10^{13}\,L_{\astrosun}$)		&$\mathbf{0.16^{+0.07}_{-0.05}}$		&$0.6^{+0.3}_{-0.2}$\\
SFR$_{\rm IR}$ ($M_{\astrosun}\,{\rm yr^{-1}}$)		&$\mathbf{237^{+97}_{-69}}$			&$862^{+403}_{-293}$\\
\cmidrule(lr){1-3}
$R_{\rm 1/2,\,CO(4-3)}$ (kpc)						&$1.31\pm0.15$		&$2.52\pm0.01$$\,^{(\ddagger)}$\\
$R_{\rm 1/2,\,3mm}$ (kpc)						&$1.1\pm0.2$			&$2.2\pm0.3$\\
\bottomrule
\end{tabular}
}
\tablefoot{The mass budgets are estimated within a given radius $r$. Where not specified, they represents the total mass of the observable emission as described in the main text. $^{(1)}${\rm The reported molecular gas masses are derived assuming a CO(4--3)-to-CO(1--0) line luminosity ratio of $r_{41}=0.87$ and a CO-to-H$_{2}$ conversion factor of $\alpha_{\rm CO}=1.7\,M_{\astrosun}\,({\rm K\,km\,s^{-1}\,pc^{2}})^{-1}$ \citep{CarilliWalter2013, MontoyaArroyave+2023}}. The uncertainties do not account for systematics on $L'_{\rm CO(4-3)}$-to-$M_{\rm H2}$ conversion factors (see Sect.~\ref{sect:md-mh2}). For the quasar host we report values of the total molecular gas mass, that accounting only for the CO(4--3) luminosity of the systematic component (``sys''), and the one excluding the blueside of the broad component (``bulk'', see Sect.~\ref{sect:md-mh2}). $^{(2)}$Mean and the standard deviation of the Gaussian prior employed in the fit of the dust SED. $^{(3)}$The FIR luminosity within the rest-frame wavelength range $8-1000\,{\rm \mu m}$. The fiducial characteristic values for the dust properties are reported in boldface. $\,^{(\dagger)}$The dynamical mass are estimated assuming dispersion-dominated system (``disp'') and rotationally-supported galaxy disk (``rot''). $\,^{(\ddagger)}$Half-light radius as derived from {\sc GalPak}$^{\rm 3D}$ modeling (see Sect.~\ref{sssect:kin-model}). }
\end{table}

\subsection{Stellar mass estimation of MQN01-QC from JWST/NIRCam images}\label{ssect:mstar}
We derive an estimate of the stellar mass ($M_{\star}$) of MQN01-QC galaxy by using the available JWST/NIRCam {\rm images with QSO PSF and background subtracted (denoted as ``iter 2'' in Sect.~\ref{ssect:qso-psf})}. We acquired photometric measurements of the MQN01-QC emission within a circular aperture $0\rlap{.}{\arcsec}3$ (corresponding to $R_{1/2,\,\rm 3mm}$ of the source; see Sect~\ref{sect:md-mh2}, and Fig.~\ref{fig:det-maps}), {\rm which is substantially larger than the resolution of NIRCam images (i.e., $0\rlap{.}{\arcsec}05$ and $0\rlap{.}{\arcsec}11$ for 1.5- and 3.2-${\rm \mu m}$ NIRCam filter, respectively). As a result, we obtained $f_{\rm 1.5\,\mu m} = 0.159\pm0.009\,{\rm \mu Jy}$, and $f_{\rm3.2\,\mu m} = 1.57\pm0.02\,{\rm \mu Jy}$} \footnote{We note that uncertainties on these flux measurements are derived from a corresponding empty aperture and do not take into account any potential systematic errors.}. Interestingly, we note that the galaxy appears to be an exceptionally red object, more than $\sim 10\times$ brighter in the $3.2\,{\rm \mu m}$ than in the $1.5\,{\rm\mu m}$ filter, reminiscent of some of the reddest galaxies recently discovered at similar or higher redshifts \citep[see, e.g.,][]{Xiao+2023}. We estimated its stellar mass using an empirical calibration between the mass-to-light ratio at $3.2\,{\rm \mu m}$ and the galaxy flux ratio $f_{\rm3.2\,\mu m}/f_{\rm 1.5\,\mu m}$ (or, equivalently, the color index $M_{F150W2}-M_{F322W2}$). {\rm Below we briefly summarize our method: (i) we collected the sample of all star-forming galaxies in MQN01 protocluster field with available stellar mass estimates derived through SED fitting as described in \citet{Galbiati+2025}; (ii) we measured the fluxes of these galaxies in the 1.5- and 3.2-${\rm \mu m}$ JWST/NIRCam images using identical apertures after matching the image PSF; (iii) we derived the best-fit model calibration by minimizing the scatter in the log-linear relation $\log\,[(M_{\star}/M_{\astrosun})/(f_{\rm 3.2\,\mu m}/{\rm mJy})] = a\,\times\,\log\,(f_{\rm3.2\,\mu m}/f_{\rm 1.5\,\mu m}) + b$. Through this method, we obtained the best calibration for $a=1.20$ and $b=12.56$, and found that the galaxy stellar mass can be predicted from the $f_{\rm3.2\,\mu m}/f_{\rm 1.5\,\mu m}$ flux ratio to an accuracy of $0.05\,{\rm dex}$ \citep[see also][for a similar approach]{Kouroumpatzakis+2023}. By applying this calibration for the observed flux ratio of $\log\,(f_{\rm3.2\,\mu m}/f_{\rm 1.5\,\mu m})=0.97\pm0.03$ measured for MQN01-QC, we derived a galaxy stellar mass (within the inner $<0\rlap{.}{\arcsec}3$ or $<2.3\,{\rm kpc}$) of $\log\,(M_{\star}^{\rm QC}/M_{\astrosun})=10.96\pm0.06\,(\pm\,0.2)$, where we also report in parenthesis the systematic uncertainty associated to the galaxy SED fitting assumptions as quantified by \citet{Pacifici+2023}.} We report this measurement in Table~\ref{tbl:budgets}. This value of the galaxy stellar mass, in combination with the estimated ${\rm SFR_{IR}}$ (see Sect.~\ref{sect:md-mh2}) place the MQN01-QC above to the star-forming main sequence at $z\approx3$ \citep{Popesso+2023}, thus suggesting that the galaxy is in a dust-obscured starburst phase\footnote{Here we assumed that any contribution from the unobscured star formation to the total SFR is negligible due to the very dusty nature of the MQN01 galaxy.}. Nevertheless, we note that this estimate is only accurate in terms of order-to-magnitude, considering the photometric systematic uncertainties originating from the quasar light subtraction and the lack of flux measurements from other filters.

\section{Kinematical analysis}\label{sect:kinematics}
In this work, we perform a full kinematical modeling of the MQN01-QC galaxy. In what follows, we describe the details of our analysis and provide an estimate of the galaxy dynamical mass. On the other hand, despite a similar analysis on the quasar host galaxy remain challenging without higher angular resolution ALMA observations, we provide virial estimates of its dynamical mass.

\subsection{Rotating disk modeling of the MQN01-QC}
We measured the geometry and kinematics of the MQN01-QC disk by employing the forward modeling technique through $^{\rm 3D}${\sc Barolo} \citep[3D-Based Analysis of Rotating Objects via Line Observations;][]{DiTeodoro+2015}\footnote{\rm The code is publicly available at \url{https://editeodoro.github.io/Bbarolo/}}, in combination with {\sc CANNUBI} \citep[Center and Angles via Numerical Nous Using Bayesian Inference;][]{Roman-Oliveira+2023}\footnote{\rm See \url{https://www.filippofraternali.com/cannubi} for full details.}. Below we summarize the main characteristics of the codes. $^{\rm 3D}${\sc Barolo} creates 3D realizations of a tilted-ring model \citep{Rogstad+1974}. Through this procedure, a galaxy is subdivided in concentric rings, each with its geometry and kinematics determined by the center $(x_{0}, y_{0})$, the inclination ($i$) and position angle (PA), the vertical thickness ($z_{0}$), the systemic velocity ($V_{\rm sys}$), the rotation velocity ($V_{\rm rot}$), the radial velocity in the disk midplane ($V_{\rm rad}$), and the velocity dispersion ($\sigma_{\rm disp}$). Under thin disk assumption ($z_{0}\approx0$) with negligible radial motions ($V_{\rm rad}\approx 0$), the line-of-sight velocity at a given galactocentric radius ($R$) can be expressed as:

\eq{
V_{\rm los} = V_{\rm sys}+V_{\rm rot}(R)\cos\phi\sin i,\label{eq:vel_los}}

where $\phi$ is the azimuthal angle in the disk plane. From the disk model, $^{\rm 3D}${\sc Barolo} creates a mock datacube that is then convolved with the PSF of the observations. This accounts for the beam smearing effect \citep{Bosma1978, Begeman1987, Epinat+2010}, which is critical for a robust measurement of the intrinsic rotational velocity of disk galaxies, especially when observed at moderate or low spatial resolution. Finally, it performs the fit to the real data by minimizing the residuals.

The galaxy kinematical model of $^{\rm 3D}${\sc Barolo} is determined by a set of free parameters, including those for the geometry of the disk. However, the limited S/N and angular resolution of the current data hamper a simultaneous determination of all the free parameters. For this reason, we employed {\sc CANNUBI} for a robust beforehand determination of the geometry of the galaxy disk. The code makes use of $^{\rm 3D}${\sc Barolo} to generate mock datacube and total flux maps obtained from model with different parameters, hence it minimizes residuals between the data and the models via a Bayesian MCMC algorithm. In this way, {\sc CANNUBI} is able to retrieve robust best-fit estimates and uncertainties of the disk geometry, such as the coordinates of the center, the radial extent, the thickness, the position angle, and the inclination. In particular, the disk inclination angle is crucial for accurately estimating the intrinsic rotation velocity, as it strongly influences the line-of-sight velocity correction (see Eq.~\ref{eq:vel_los}). In this work, we assume a razor-thin disk model, and therefore that the observed galaxy geometry is entirely due to the projection of an intrinsically round disk on the sky plane. Under such assumptions, the inclination is directly determined by $\sin^{2}i=1-(b_{\rm min}/a_{\rm maj})^{2}$, where $a_{\rm maj}$ and $b_{\rm min}$ are the semi-major, and semi-minor axes of the {\rm intrinsic galaxy disk model}, respectively. In the next sections, we describe our methods and findings.

       \begin{figure*}[!ht]
   	\centering
   	\resizebox{\hsize}{!}{
		\includegraphics{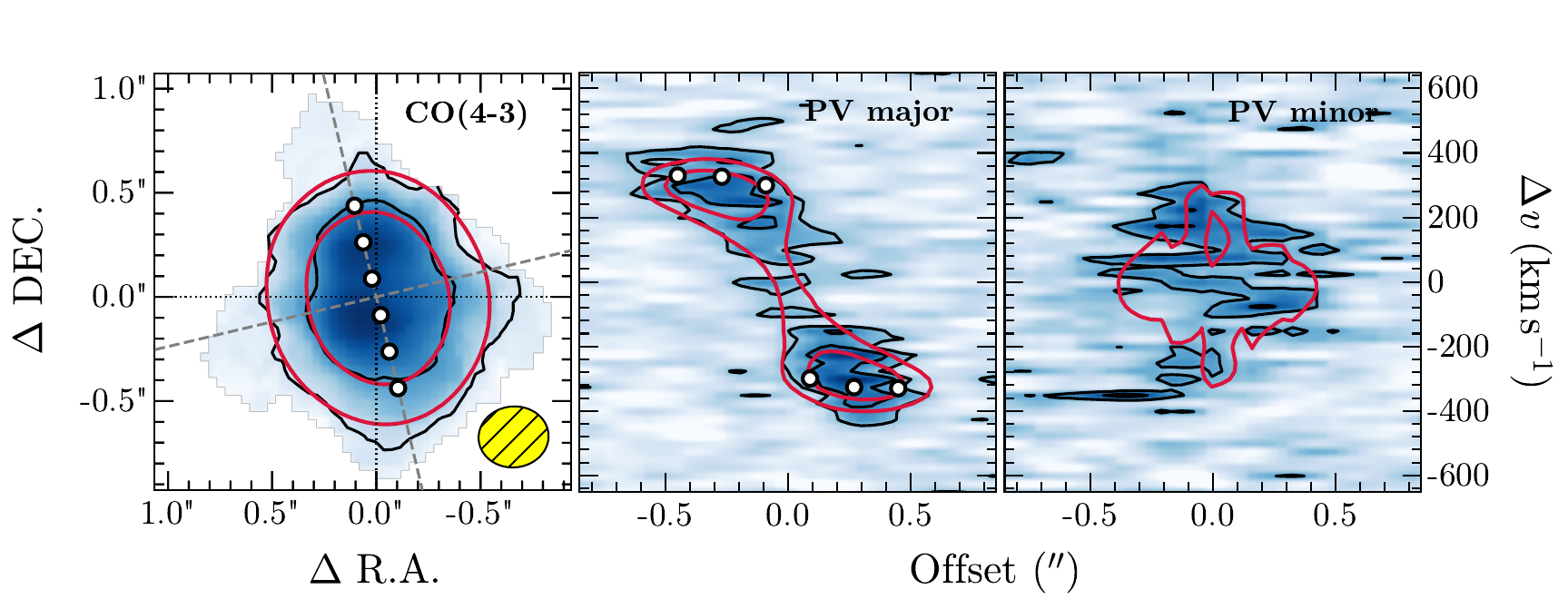}}
       \caption{Geometrical and kinematical modeling results as obtained with {\sc CANNUBI} and $^{\rm 3D}${\sc Barolo}. {\it Left panel:} the observed CO(4--3) line-velocity integrated map overlaid with the best-fit model. The black and red contours show the $[4,16]\sigma$ isophotes of the data and the model, respectively, where $\sigma$ denotes the noise RMS {\rm in the flux map}. The gray dashed lines indicate the best-fit location of the major and minor axes of the disk. The white circles show the spacing between the model rings. The ALMA synthesized beam is reported in the bottom right corner. {\it Central and right panels:} PVDs along the best-fit major and minor axis of the model. The black and red contours show the $[2,4]\sigma$ isophotes of the data and the best-fit rotating thin-disk model, respectively, {\rm where $\sigma$ is the noise RMS per velocity channel.}}
       \label{fig:modeling}
    \end{figure*}

\subsubsection{Geometrical parameter derivation}
We run {\sc CANNUBI} on the continuum-subtracted datacube of MQN01-QC by minimizing the sum of $\abs{{\rm data}-{\rm model}}$ using three annuli ({\tt NRADII=3}). These are sufficient to cover the entire emitting line region ($\sim 1\arcsec$ extension) given the resolution of the data ($\simeq0\rlap{.}{\arcsec}3$). We set {\tt errorType=``marasco''}, which calculates the likelihood taking into account the asymmetries in the total intensity map \citep[i.e., the moment 0th; see][]{Marasco+2019}. For the masking, we first applied a smoothing to a $1.5\times$ lower resolution and then scan the smoothed data for sources building the mask with {\tt MASK=``SMOOTH\&SEARCH''} task adopting {\tt FACTOR=1.5}, and {\tt SNRCUT=3}, {\tt GROWTHCUT=2} as primary and secondary S/N threshold, respectively. In Table~\ref{tbl:morphkin-pars}, we report the best fit values and uncertainties obtained from the posterior probability distributions of the free parameters. We also show the posteriors in Appendix~\ref{app:posteriors}. In Fig.~\ref{fig:modeling}, we show the best fit model overlaid with the data. 

\begin{table}[!t]
\def\arraystretch{1.15}
\caption{Geometrical and kinematical best-fit parameters of the MQN01-QC galaxy disk.}  
\label{tbl:morphkin-pars}      
\centering 
\resizebox{0.9\hsize}{!}{
\begin{tabular}{ll} 
\toprule\toprule
Geometrical parameters  from {\sc CANNUBI}&\\
\cmidrule(lr){1-2}
\multirow{2}{*}{Radial Extent} 	&	$4.1\pm0.4\,{\rm kpc}$$\,^{(1)}$\\
						&	$0.54\pm0.05\,{\rm arcsec}$\\
Ring Width$\,^{(2)}$ (arcsec)	&	$0.179^{+0.018}_{-0.016}$ \\
$i$ (deg)					&	$40^{+10}_{-14}$	\\
PA (deg) 					&  	$193^{+19}_{-16}$	\\
\bottomrule
Kinematical parameters from $^{\rm 3D}${\sc Barolo}&\\
\cmidrule(lr){1-2}
$V_{\rm rot, max}$$\,^{(3)}$ (${\rm km\,s^{-1}}$)	& 	$513^{+58}_{-64}$\\
$\bar{\sigma}$$\,^{(4)}$ (${\rm km\,s^{-1}}$)		&	$46^{+12}_{-12}$\\
$V_{\rm rot, max}/\bar{\sigma}$					&	$11^{+4}_{-3}$\\
\bottomrule
Kinematical parameters from {\sc GalPak}$^{\rm 3D}$&\\
\cmidrule(lr){1-2}
 $V_{\rm rot, max}$ (${\rm km\,s^{-1}}$)			&	$531\pm4$\\
 $\sigma_{0}$$\,^{(5)}$ (${\rm km\,s^{-1}}$)				&	$54\pm 1$\\
 $V_{\rm rot, max}/\sigma_{0}$					&	$10.0\pm0.2$\\
\bottomrule
\end{tabular}
}
\tablefoot{$^{(1)}$Disk radial extension obtained from the best-fit estimate of the radial separation of the annuli assuming $z=3.2475$. $^{(2)}$The separation between rings corresponds to an oversampling of a factor of $1.6\times$ of the synthesized beam minor axis. Hence, the parameters derived for each ring are not fully independent. $\,^{(3)}$Maximum disk rotational velocity. $\,^{(4)}$Radially averaged velocity dispersion of the disk. $\,^{(5)}$Intrinsic velocity dispersion of the disk. The reported errors on kinematical quantities do not take into account that on disk geometry and are therefore underestimated.}
\end{table}
\subsubsection{Kinematical parameter derivation}\label{sssect:kin-model}
We run $^{\rm 3D}${\sc Barolo} on the continuum-subtracted datacube of MQN01-QC. We employed the same mask parameters and three annuli with fixed radial separation, inclination, and position angle corresponding to the best-fit values as derived from the morphological fit with {\sc CANNUBI}, and we left the rotational velocity ({\tt VROT}), and the velocity dispersion ({\tt VDISP}) as free parameters. We set the systemic velocity at the CO(4--3) line redshift of $z=3.2475$\footnote{This corresponds to the flux-weighted average redshift of the two Gaussian components of the CO(4--3) line profile model of MQN01-QC (see Table.~\ref{tbl:sou_flux_prop}).}. We also assumed negligible radial velocity ($V_{\rm rad}=0$), as the isovelocity curves of the CO-traced gas appear to be not significantly distorted \citep[see, e.g.,][]{Lelli+2012b,Lelli+2012a}\footnote{We note that the current S/N and angular resolution of the observations are insufficient to investigate the presence of potential radial motions. However, since the observed molecular gas kinematics is well reproduced by a purely rotating disk, we find no compelling evidence to support the presence of radial motions.}. Finally, we employed an axisymmetric disk normalization by setting {\tt NORM=``AZIM''}. With this assumption, the surface density of each ring is computed directly from the observed flux map using an azimuthal average. Although the observed CO(4--3) disk in the MQN01-QC is not perfectly axisymmetric, this option enables a robust determination of the rotational velocity in the outer rings, which is a proxy for the enclosed dynamical mass of the galaxy. Through $^{\rm 3D}${\sc Barolo}, we minimized the absolute difference between the data and the model employing $\cos^{2}\phi$ as weighting function {\tt WFUNC=2}. This choice gives prominence to regions close to the major axis, which enclose most of the information on the rotational motion. In Table~\ref{tbl:morphkin-pars}, we report the best-fit values for the maximum rotational velocity ($V_{\rm rot, max}$) and the radially-averaged velocity dispersion ($\bar{\sigma}$). We note that the reported uncertainties do not take into account those on geometry. 

{\rm In Fig.~\ref{fig:spectra} and~\ref{fig:modeling}, we compare, respectively, the line profile and the position-velocity diagrams (PVDs) along the major and minor axis of the best-fit disk model, with the observed data of the MQN01-QC galaxy.} The PVD along the major axis is well reproduced by our rotating disk model, which has a steeply rising rotation curve. The observed emission in the minor-axis PVD seems slightly asymmetric. This is most likely an effect of the low S/N, rather than a signature of radial motions. The vertical extend of the emission is captured by our model, which indicates that the velocity dispersion is robustly determined. Overall, the kinematics of the MQN01-QC is broadly reproduced by the rotating thin-disk model with no strong evidence of non-circular motions. This is also evident from the good match between data and model in the datacube channel maps (see Appendix~\ref{app:chan-maps}). Following a common approach used in literature \citep[e.g.,][]{ForsterSchreiber+2018, Wisnioski+2019, Lelli+2023, Rizzo+2020, Rizzo+2024, Kohandel+2024}, we quantify the level of rotational support by computing the rotation-to-dispersion velocity ratio and obtained $V_{\rm rot, max}/\bar{\sigma}=11^{+4}_{-3}$. This implies a high degree of rotational support for MQN01-QC.

We compare our findings with that obtained with a different kinematical analysis of the MQN01-QC galaxy disk using the parametric code {\sc GalPak}$^{\rm 3D}$ \footnote{The code is publicly available at \url{https://galpak3d.univ-lyon1.fr/doc/index.htm}.}. The details about this code are described in \citet{Bouche+2015}. Briefly, {\sc GalPak}$^{\rm 3D}$ is a Bayesian parametric modeling tool which makes use of reverse deconvolution method and employs MCMC procedure to recover the intrinsic geometrical and kinematical properties of a galaxy. The code is optimized for low S/N and angular resolution observations. We modeled the MQN01-QC datacube with {\sc GalPak}$^{\rm 3D}$ by adopting a Gaussian profile for both the radial and the vertical flux distribution of the disk. We employed an arctangent model as the intrinsic rotation velocity curve of the form $V_{\rm rot}(R)=V_{\rm max}(2/\pi)\arctan(R/R_{\rm t})$. For the fit, we fixed the disk geometrical center, position angle, and inclination to the those obtained with {\sc CANNUBI}. We left as free parameters the maximum rotation velocity ($V_{\rm max}$), the turnover radius ($R_{\rm t}$), the intrinsic velocity dispersion ($\sigma_{0}$), and the systemic velocity of the disk, as well as its half-light radius ($R_{1/2}$), and the flux profile normalization. As a result, we find that the kinematical parameters are consistent within the uncertainties with those retrieved with $^{\rm 3D}${\sc Barolo} (see Table~\ref{tbl:budgets}, and \ref{tbl:morphkin-pars}).

      \begin{figure*}[!ht]
   	\centering
   	\resizebox{0.9\hsize}{!}{
		\includegraphics{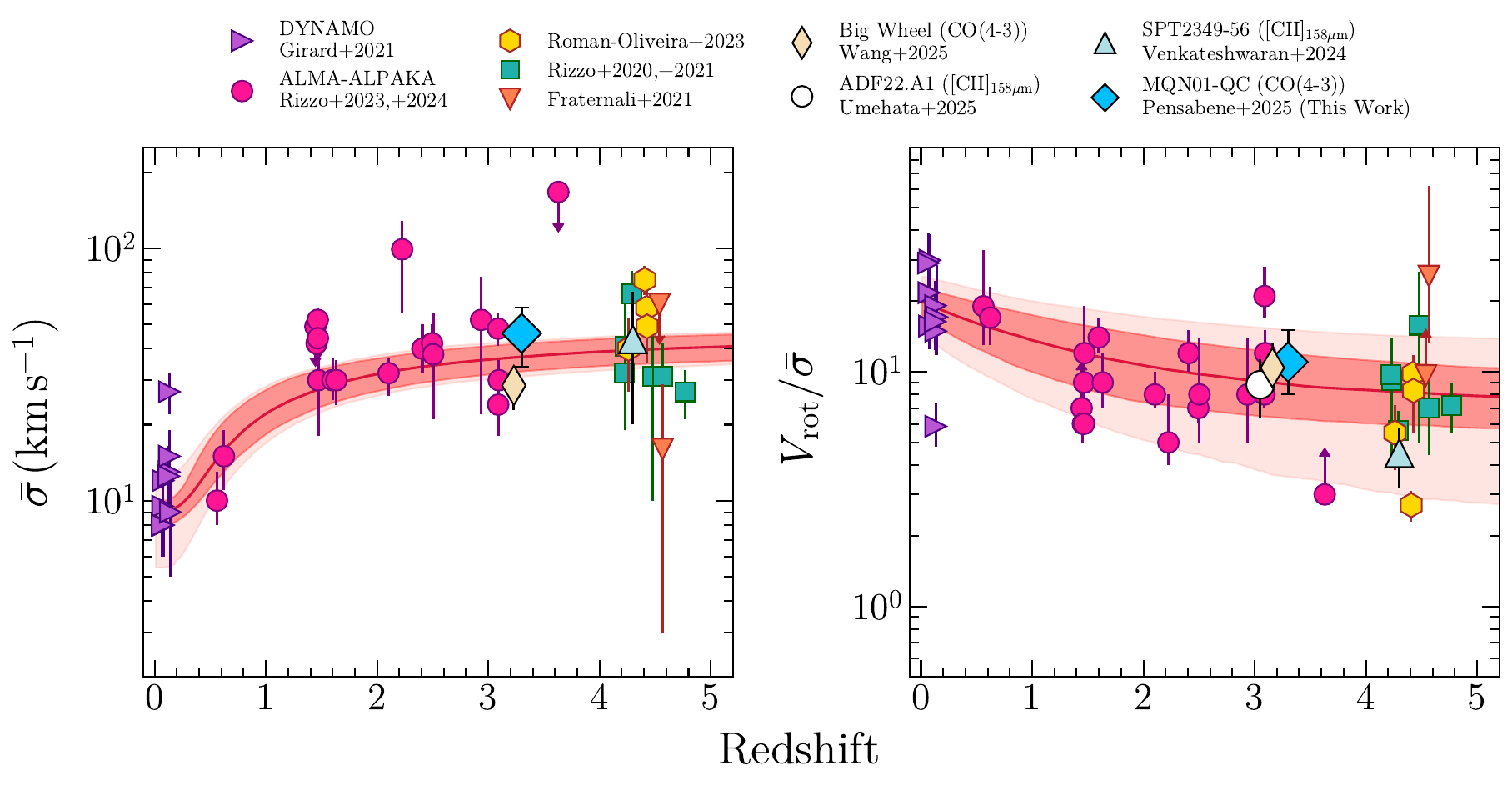}}
       \caption{Redshift distribution of the mean velocity dispersion ({\it left panel}), and the degree of rotational support ({\it right panel}) for a sample of main-sequence and starburst galaxy disks. Datapoints are taken from \citet{Rizzo+2020, Rizzo+2021, Rizzo+2023, Rizzo+2024, Fraternali+2021, Girard+2021, Roman-Oliveira+2023} who analyzed galaxy cold gas kinematics traced by CO emission lines. The $V_{\rm rot}/\bar{\sigma}$ values for giant spiral galaxies in MQN01 and SSA22 protocluster from \citet{Wang+2025} (CO(4--3)), \citet{Umehata+2025} (\cii{}) are also reported by the beige diamond and the white circle, respectively, and slightly shifted horizontally for clarity. {\rm The mean velocity dispersion and rotational support of galaxies in SPT 2349-56 protocluster core as traced by the \cii{} line emission are also reported with the light blue triangle \citep{Venkateshwaran+2024}}. Our measurements for the MQN01-QC galaxy are indicated by the blue diamonds. The red solid line is the best fit model derived in \citet{Rizzo+2024}. The dark and light shaded areas represent the 1-$\sigma$ confidence interval on the best-fit relation and its intrinsic scatter, respectively.}
       \label{fig:vrot-sigma}
    \end{figure*}
    
\subsection{Dynamical and virial mass estimates of galaxies}\label{ssect:dyn-mass}
We used our measurements of the disk radial extent and maximum circular velocity to estimate the dynamical mass ($M_{\rm dyn}$) of MQN01-QC galaxy. Under the assumption that the mass distribution is spherically symmetric, the dynamical mass enclosed within a radius $R$, is given by

\eq{M_{\rm dyn}(<R)\,[M_{\astrosun}]  = \frac{\varv^{2}_{\rm circ}}{G}R=2.33\times10^{5}\,\tonda{\frac{\varv_{\rm circ}}{\rm km\,s^{-1}}}^{2}\tonda{\frac{R}{\rm kpc}}\label{eq:dyn-mass},}
where $G$ is the gravitational constant, and $v_{\rm circ}$ is the circular velocity. The latter is evaluated from the rotational velocity taking into account the contribution from random motions \citep[see, e.g.,][]{Epinat+2009, Burkert+2010, Turner+2017, Iorio+2017, Lelli+2023}. Since $V_{\rm rot }/\sigma\sim 10$ for the MQN01-QC disk, this implies negligible pressure support, thus making $\varv_{\rm circ}\approx V_{\rm rot}$. We note that the approximation of spherically mass distribution is likely an oversimplification in our case, as the baryonic disk component is expected to be dominant in such inner regions of galaxies. Under our assumptions, we expect the mass measurement to be overestimated by $\sim30\%$ at most \citep[see, e.g.,][]{BinneyTremaine2008, Price+2022}. Therefore, following \citet{Neeleman+2021}, we conservatively include a $20\%$ of uncertainty toward lower mass to take into account this potential systematic. Given the aforementioned considerations, we estimated a dynamical mass of the MQN01-QC galaxy within the observed radial extent of $M_{\rm dyn}^{\rm QC}(< 4.1\,{\rm kpc}) = 2.5^{+0.6}_{-1.1}\times 10^{11}\,M_{\astrosun}$.
 
For the quasar host, we estimate the galaxy virial mass under the assumption of dispersion-dominated system through the virial relation $GM_{\rm vir} \approx k\sigma R^{2}$. Here, the constant $k$ is intended to account for the ``degree of virialization'' \citep[see, e.g.,][]{Cappellari+2006, Taylor+2010, vanderWel+2022}. Following \citet{Decarli+2018}, who estimated virial masses of a large sample of $z>6$ quasar host galaxies, we employ $k=3/2$. For $\sigma$, we use the width of the systemic component of the CO(4--3) line profile (see Table~\ref{tbl:sou_flux_prop}), in order to exclude potential non-virialized structures (blueshifted component in Fig.~\ref{fig:spectra}). To estimate the size of the quasar host galaxy, we modeled the CO(4--3) line map with a 2D Gaussian using the CASA task {\tt imfit} and found a beam-deconvolved major and minor axis FWHM of $a_{\rm maj}=0.34\pm0.04\,{\rm arcsec}$ and $b_{\rm min}=0.27\pm0.04\,{\rm arcsec}$, respectively. This corresponds to a radial extent of $R_{\rm 1/2,\,CO(4-3)}^{\rm QSO} = 1.31\pm0.15\,{\rm kpc}$ at $z=3.2510$. We report such measurement in Table~\ref{tbl:budgets}. We note that this size is consistent within the uncertainties with the half-light radius of quasar host galaxy as measured from the 3-mm continuum emission (see Sect.~\ref{sect:md-mh2}). This suggests that the distribution molecular gas and dust have comparable extension. By using these assumptions and estimates, we find $M_{\rm vir}^{\rm QSO}(< 1.3\,{\rm kpc}) = 1.3^{+0.6}_{-0.4}\times 10^{10}\,M_{\astrosun}$. 

Alternatively, in the extremely opposite case in which the bulk of the emitting gas is dominated by regular rotation, we estimate the quasar host dynamical mass through the relation $M_{\rm dyn} = G^{-1}R\,(0.75\times {\rm FWHM}/\sin i)^{2}$ \citep[see, e.g.,][]{Ho+2007, Wang+2013, Willott+2015mbh, Decarli+2018}. Similarly to above, we use as FWHM the linewidth of the quasar CO(4--3) systemic component. Finally, we estimated the galaxy inclination angle through the axial ratio as $i = \cos^{-1}(b_{\rm min}/a_{\rm maj}) = 40\pm 10\, {\rm deg}$, and found $M_{\rm dyn}^{\rm QSO}(< 1.3\,{\rm kpc})=0.7^{+0.8}_{-0.3}\times 10^{11}\,M_{\astrosun}$. 

We note that, given the limited angular resolution of the observations, the above virial mass estimates suffer from large uncertainties and should be taken with caution. However, they can be used as a first guess of the dynamical mass of the quasar host galaxy \citep[see][for further discussion]{Pensabene+2020, Neeleman+2021}. Taken at face values, our estimates suggest that the quasar host galaxy might be comparably or at most $\sim 10\times$ less massive than the closely-separated MQN01-QC galaxy. The observed system may constitute a early-stage galaxy merger. In this scenario, if the quasar host dynamical mass matches the estimate for a dispersion dominated system, it would likely act as the satellite of its companion galaxy. We further discuss this hypothesis in Sect.~\ref{ssect:dynamics}.

\section{Discussion}\label{sect:discussion}
\subsection{Comparison with other cold disks across cosmic time}
Numerous lines of evidence shows that massive star-forming galaxies detected with ALMA can be settled in rotationally supported disks even in the early Universe \citep[e.g.,][]{Neeleman+2020, Rizzo+2020, Rizzo+2023, Rizzo+2024, Fraternali+2021, Lelli+2021, Lelli+2023, Ginolfi+2022, Roman-Oliveira+2024, Rowland+2024}, rather than being characterized by irregular morphologies and strong turbulent motions, as expected by some theoretical models as a consequence of interaction or intense gas accretion \citep[see][]{Hopkins+2014, Schaye+2015, Vogelsberger+2020}. {\rm Our kinematical analysis shows that, despite residing in the immediate proximity of a hyperluminous quasar within one of the most significant galaxy overdensity identified at cosmic noon (see Paper I; \citealt{Galbiati+2025, Travascio+2025}), MQN01-QC exhibits a high level of rotational support ($V_{\rm rot}/\bar{\sigma} \approx 11$) comparable to that of other dynamically cold rotating disks at similar redshifts \citep[see, e.g.,][]{Rizzo+2024}}. Our finding suggests that cold disks are able to survive even in such a hostile environment, where galaxy interactions, mergers, and ram-pressure stripping may disrupt ordered gas kinematics. 

     \begin{figure}[!t]
   	\centering
   	\resizebox{\hsize}{!}{
		\includegraphics{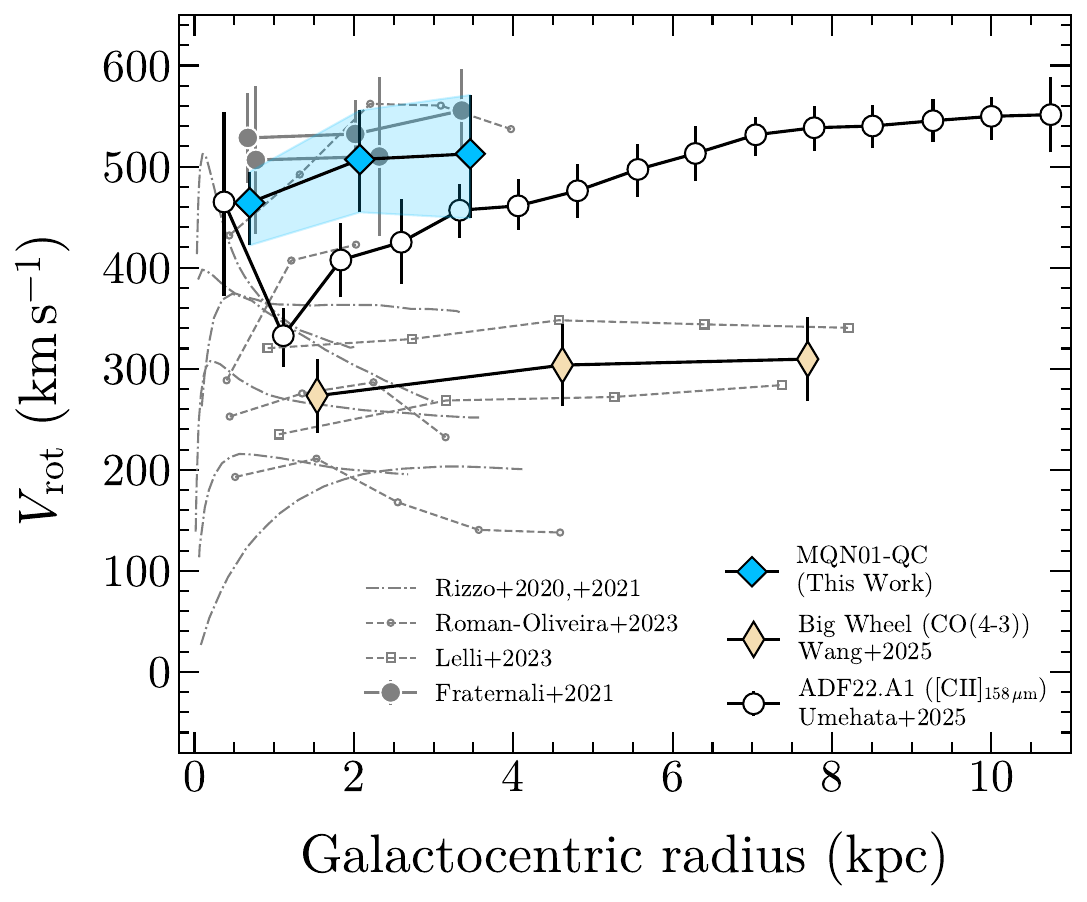}}
       \caption{Velocity curves of $z\apprge1.5$ main-sequence and starburst galaxy disks from \citet[][]{Rizzo+2020, Rizzo+2021, Fraternali+2021, Lelli+2023, Roman-Oliveira+2023}, and those of giant spiral galaxies discovered in $z\sim 3$ protoclusters (ADF22.A1, \citealt{Umehata+2025}, and Big Wheel, \citealt{Wang+2025}). The MQN01-QC velocity curve as derived in this work is reported by the blue diamonds.}
       \label{fig:vrot-curve}
    \end{figure}
    
In Fig.~\ref{fig:vrot-sigma}, we compare the velocity dispersion and the rotational support measurements of the MQN01-QC galaxy disk with the empirical relations from \citet{Rizzo+2024}. The latter has been derived from samples of main-sequence and starburst galaxies observed primarily through CO lines. We found that our kinematical measurements for MQN01-QC galaxy are in full agreement with the observed evolution of turbulence across cosmic time. \citet{Rizzo+2024} show that the observed evolutionary trends can be explained by models for turbulence driven by stellar feedback \citep[see also][]{Bacchini+2020, Fraternali+2021}. We apply this model on the MQN01-QC to estimate the efficiency of the supernovae (SNe) in transferring kinetic energy to the molecular gas. We employ equation 10 in \citet{Rizzo+2024}, and the $M_{\rm H2}$ and SFR measurements from Table~\ref{tbl:budgets}. We further adopt a SN rate of $\eta_{\rm SN}=1.3\times10^{-2}\,M_{\astrosun}^{-1}$, and a disk scale height of $h=0.2\,{\rm kpc}$ \citep[see][]{Bacchini+2024}. Under these assumptions, we found $\epsilon_{\rm SN} \simeq 0.05$, implying that only $5\%$ of the kinetic energy from SN feedback is needed to drive the observed turbulence in MQN01-QC disk. This result is fully in line with the typical values measured from cold gas kinematics in dusty star-forming galaxies (DSFGs) at $z\apprge3$ \citep{Rizzo+2020,Rizzo+2021,Roman-Oliveira+2024}. 

In Fig.~\ref{fig:vrot-curve}, we compare the rotation curve of the MQN01-QC galaxy and other DSFG disks at $z\sim3-4$ from the literature, including also the velocity curves derived for two giant spiral galaxies recently discovered in protocluster environments at $z\sim3$ \citep[][]{Umehata+2025, Wang+2025}. The MQN01-QC galaxy disk reaches a maximum value of $\sim500\,{\rm km\,s^{-1}}$ at $\simeq 4\,{\rm kpc}$ which is much higher than the typical $V_{\rm rot, max}$ measured in DSFGs at $z\sim3-4$ \citep[see, e.g.,][]{Rizzo+2020, Rizzo+2021, Lelli+2023, Roman-Oliveira+2023, Amvrosiadis+2025}. Interestingly, the MQN01-QC rotation curve resembles those of two $z\simeq 4.5$ starburst galaxies reported by \citet{Fraternali+2021}. Notably, \citealt{Fraternali+2021} found that, assuming all their gas is converted to stars, these systems overlap with the massive early-type galaxies (ETGs) at $z=0$, on the ETG-analogue of the stellar-mass Tully-Fisher \citep{Tully+1977, Davis+2016}. In other words, there is a dynamical evidence for an evolutionary link between massive high-redshift starbursts and ETGs. We do the same experiment by adopting $M_{\rm dyn}^{\rm QC}=2.5\times 10^{11}\,M_{\astrosun}$ (see Table~\ref{tbl:budgets}) as an upper limit on the baryonic mass budget of the MQN01-QC galaxy, and find that MQN01-QC galaxy falls within the same parameter space of local ETGs. This suggests that MQN01-QC could indeed represent one of the progenitors of today's most massive ellipticals.

       \begin{figure}[!t]
   	\centering
   	\resizebox{0.9\hsize}{!}{
		\includegraphics{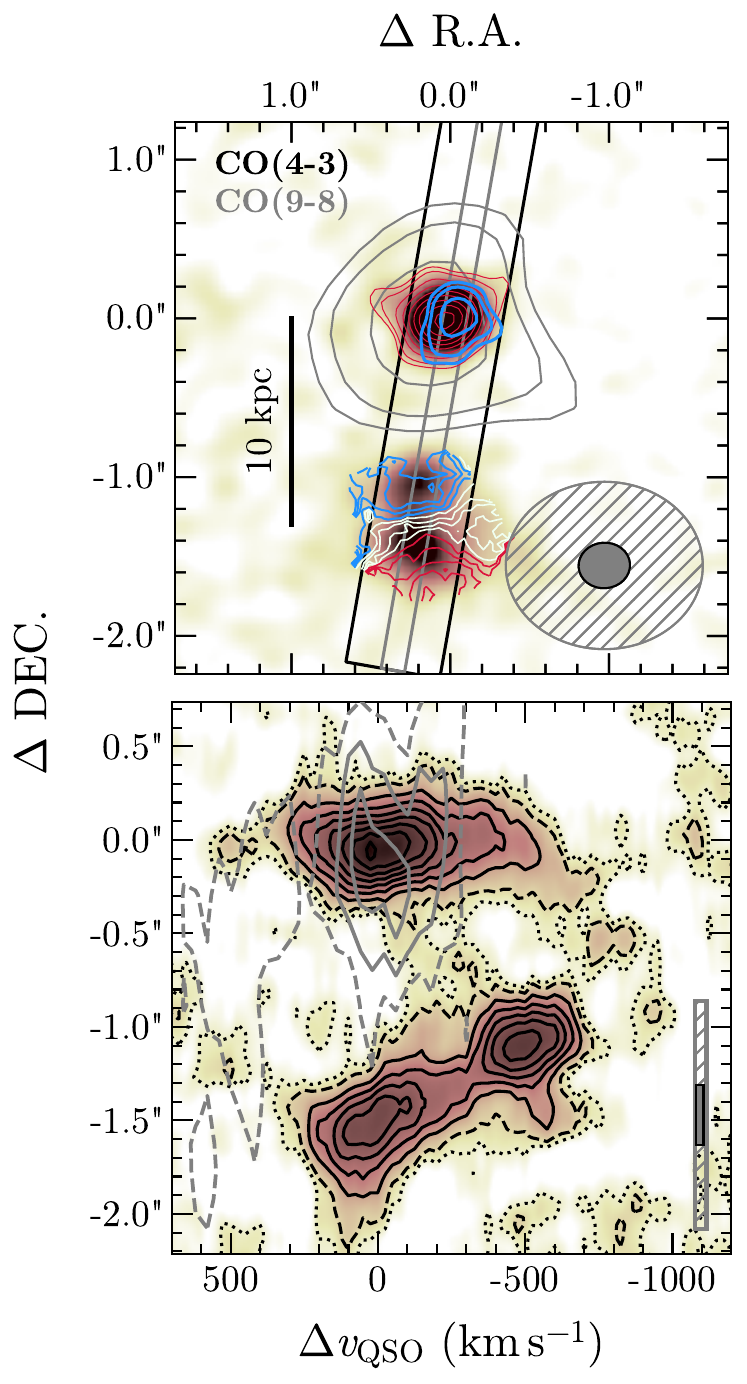}}
       \caption{Search for gas emission around galaxies. {\it Top panel:} The background image and gray contours represent the total CO(4--3) and CO(9--8) intensity map, respectively. These are obtained within $\Delta\varv_{\rm QSO}=[-700;+400]\,{\rm km\,s^{-1}}$. The red and blue contours on the quasar host galaxy show residual line maps after subtracting from the CO(4--3) total flux map, the blueshifted $[-900,-300]\,{\rm km\,s^{-1}}$, and the systemic $[-300,+400]\,{\rm km\,s^{-1}}$ image component, respectively. The contours correspond to $[3,2n]\sigma$, with $n>1$ is an integer number, and $\sigma$ denotes the noise RMS. The contours on the MQN01-QC shows instead the isovelocity curves (blue: $[-500,-350]\,{\rm km\,s^{-1}}$; white: $[-300,-150]\,{\rm km\,s^{-1}}$; red: $[-100, 50]\,{\rm km\,s^{-1}}$). The synthesized beam of the band 6 and 3 observations are reported with the gray hatched and filled ellipse at the bottom right corner, respectively. {\it Bottom panel:} PVD extracted along the black ($0\rlap{.}{\arcsec}6$-wide), and the gray ($0\rlap{.}{\arcsec}15$-wide) slit, for the band 3 and 6 datacube, respectively (see top panel). The background image and the gray contours report the continuum-subtracted band 3 and 6 data, respectively. The resolution of the each dataset is show with rectangles at the bottom right corner. The dotted dashed and solid contours represent $1\sigma$, $2\sigma$ and $2n\sigma$ level, respectively.}
       \label{fig:res-maps}
    \end{figure}
    
\subsection{Origin of the blueshifted emission of the quasar host galaxy}\label{ssect:dynamics}
The observed blueshifted component detected in the CO(4--3) line profile of the quasar host galaxy (see Fig.~\ref{fig:spectra}), may be explained by two main possible scenarios: 1) a tidal tail generated by the interaction between the quasar host and its MQN01-QC companion galaxy; 2) a molecular gas outflow driven by the quasar. In this section we test these scenarios to determine which one best matches our observations.

We firstly inspect the datacube to search for any evidence of gas tails or ``bridges'' that might be indicative of interactions between galaxies and/or with the surrounding medium. In Fig.~\ref{fig:res-maps}, we report the PVD extracted from a slit connecting the quasar host and the MQN01-QC galaxy. Through this method, we detect -- albeit tentatively (${\rm S/N\apprge 1}$) -- blueshifted gas emission traced by the CO(4--3) line up to line-of-sight velocities of approximately $-900\,{\rm km\,s^{-1}}$, which appears also stretched toward the approaching side of the MQN01-QC galaxy. We also note that we do not detect a similar feature in the CO(9--8) line emission of the quasar host galaxy which seems to be mostly associated to the CO(4--3) systemic component. This evidence, alongside the close projected distance between between the galaxies, might support the idea that tidal interactions are actively occurring in this system. The high degree of rotational support of the MQN01-QC disk, and the tentative dynamical mass estimates (see Sect~\ref{ssect:dyn-mass}), indicate that this system may constitute a minor merger in progress where the MQN01-QC would represent the most massive galaxy (i.e, the primary) of the system which is inducing strong gravitational disturbances on the quasar host (i.e., the satellite). 

To further test this hypothesis we compute the tidal radius of the MQN01-QC galaxy ($R_{\rm J}$) and that of the quasar host ($r_{\rm J}$), under the hypotheses of spherically symmetric gravitational potentials and that galaxies are moving on circular orbits around the center of mass of the two-body system\footnote{Following \citet{BinneyTremaine2008}, we computed the tidal radius of a satellite of mass $m$ in the gravitational potential of a point-like mass $M$ by finding numerically the saddle points of the effective two-body gravitational potential as
\eq{\frac{M}{(R_{0}-r_{\rm J})^{2}}-\frac{m}{r_{\rm J}^{2}}-\frac{M+m}{R_{0}^{3}}\tonda{\frac{MR_{0}}{M+m}-r_{\rm J}}=0,}
where $M$ is mass of the primary, $R_{0}$ is the physical separation between the two bodies, and $r_{\rm J}$ is the Jacobi radius (or tidal radius) of the satellite, that is the radius beyond which tidal interaction are expected to be dominant. The tidal radius of the primary can be found similarly by inverting $M$ with $m$ in the above equation or via $R_{\rm J}=R_{0}-r_{\rm J}$. We note that in our computation we consider the reference frame of the two body system as galaxies may have their own orbits in the protocluster potential.}. We further assume a conservative (i.e., lower limit) real-space physical separation between the galaxies of $10\,{\rm kpc}$, and the (virial) dynamical masses derived in Sect.~\ref{ssect:dyn-mass} as estimates of the total mass of the individual galaxies. With this set of assumptions, and taking into account the possible range for the dynamical mass of the quasar host, we found $R_{\rm J}\simeq 7.8-5.5\,{\rm kpc}$, and $r_{\rm J}\simeq 2.1-4.5\,{\rm kpc}$. Conversely, we may ask which should be the minimum mass of the quasar host in order to induce strong tidal disturbances on the MQN01-QC galaxy disk. To this purpose, we employ our dynamical estimate of the mass of MQN01-QC, and its radial extent as $r_{\rm J}$. This yields a minimum mass of the quasar host of $\simeq 5\times 10^{11}\,M_{\astrosun}$, which is larger than the our virial mass estimates. Taken at face values, the above estimates indicate that tidal effects are expected to be negligible in this system. This is in contrast with our intuitions based on Fig.~\ref{fig:res-maps}. However, we stress that these arguments should be taken with caution since they rely on highly simplifying assumptions which are difficult to verify. Furthermore, an accurate dynamical analysis is limited by projection effects and lack of information regarding the total mass (dark matter + baryons) of the individual galaxies. Indeed, the galaxy mass estimates we derived in Sect.~\ref{ssect:dyn-mass} do not account for the remaining mass budget distributed at larger distances outside the radius probed by the detected CO(4--3) emission. In conclusion, our analysis shows that with the current available information, we cannot completely rule out the tidal stripping scenario,  albeit the latter is not supported by our simplified calculation discussed above. 

We now consider the outflow scenario. The CO(4--3) blue wing may indeed trace outflowing gas from the front side of the quasar host. This interpretation is also supported by the observation of a broad blueshifted [OIII]$\lambda5007$ line ($FWHM\sim1500\,{\rm km\,s^{-1}}$, and velocity shift $\sim-500\,{\rm km\,s^{-1}}$) in the NIR spectrum of the quasar as found by \citet[][see their figure A.3]{Deconto-Machado+2023}, which would possibly imply the presence of a ionized gas outflow. The velocity shift ($\sim -300 \,{\rm km\,s^{-1}}$) and FWHM ($\sim 700\,{\rm km\,s^{-1}}$) of the broad component that we revealed in the CO(4--3) line profile of the quasar host, is consistent with the typical velocities of molecular outflows observed in nearby AGN \citep[see, e.g.,][]{Feruglio+2010, Veilleux+2013, Cicone+2012, Cicone+2014, Bischetti+2019b, Fluetsch+2019}, and high-redshift quasars \citep[see, e.g.,][]{Polletta+2011, Feruglio+2017, Brusa+2018, Bischetti+2019a, Bischetti+2024, Vayner+2021}. Following these previous works, we compute the molecular gas outflow rate via $\dot{M}_{\rm H2} = (M_{\rm H2}^{\rm out} v_{\rm out})/R_{\rm out}$ \citep[see][]{Rupke+2005}, where $M_{\rm H2}^{\rm out}$, $v_{\rm out}$,  and $R_{\rm out}$ are the molecular gas mass, the maximum velocity of the outflow, and its spatial extension. {\rm To compute the molecular gas mass within the outflow we adopted a CO(4--3)-to-CO(1--0) line luminosity ratio of $r_{41} = 0.87$ \citep{CarilliWalter2013}, and a conservative $\alpha_{\rm CO} = 0.5\,M_{\astrosun}\,({\rm K\,km\,s^{-1}\,pc^{2}})^{-1}$. The latter CO-to-H$_{2}$ conversion factor has been measured in the molecular outflow of M82 and is widely adopted in the literature to compute the molecular mass of outflowing gas in high redshift quasar host galaxies \citep{Weiss+2001, Feruglio+2010, Feruglio+2017, Cicone+2014}.} 
    
We consider only the blueside of the broad component at velocities of $<-300\,{\rm km\,s^{-1}}$ as the CO emitting gas associated to the outflow. We therefore take into account only half of broad line luminosity. We further adopt $v_{\rm out}\simeq770\,{\rm km\,s^{-1}}$ as the 95th percentile of the cumulative velocity distribution of the broad line component, and assume $R_{\rm out}\equiv R_{\rm 1/2,\,CO(4-3)}^{\rm QSO}$ as the outflow extension (see Sect.~\ref{ssect:dyn-mass}, and Table~\ref{tbl:budgets}). Under these hypotheses, we obtain an estimate of the molecular gas outflow rate of $\dot{M}_{\rm H2} \simeq 1.7\times10^{3}\,(\alpha_{\rm CO}/0.5)\,M_{\astrosun}\,{\rm yr^{-1}}$. This value places the outflow along the empirical relation for AGN winds \citep{Fiore+2017, Bischetti+2019b} considering that the estimated quasar bolometric luminosity is $L_{\rm Bol}=1.8\times10^{48}\,{\rm erg\,s^{-1}}$ \citet{Deconto-Machado+2023}. The depletion time scale associated to this outflow -- i.e., the timescale needed to clear out the whole gas content of the galaxy --  would be $\tau_{\rm dep}=M_{\rm H2}/\dot{M}_{\rm H2} \simeq 20\,{\rm Myr}$, when accounting for the bulk of the molecular gas within the quasar host (see Sect.~\ref{sect:md-mh2}). This value is $\sim 10\times$ shorter than the time needed for the molecular gas to be converted into stars ($\tau_{\rm SF} = M_{\rm H2}/{\rm SFR_{IR}}\simeq 150\,{\rm Myr}$), implying that the outflow is potentially able to significantly affect the evolution of the quasar host galaxy, quenching the star formation in a relatively short timescale. Similar evidence has been also reported in other works investigating molecular outflows in AGN \citep[see, e.g.,][]{Cicone+2014, Veilleux+2017, Brusa+2018, Herrera-Camus+2019}. Overall, the outflow scenario is a viable explanation to the observed broad blueshifted component observed in the quasar CO(4--3) line profile, but we cannot rule out any of the two alternative scenarios (i.e., merger or outflow) with the available information. 

Finally, we note that a molecular gas inflow from a low-mass satellite galaxy, or along one or more cosmic web filaments, or from the circumgalactic medium \citep[see, e.g.,][]{Emonts+2023}, would potentially explain the observed blueshifted CO(4--3) emission. This latter scenarios are however difficult to test with the current data. Deeper and higher angular resolution observations in the (sub-)mm band with ALMA targeting tracers of the cold atomic/molecular gas, and NIR mapping with integral field spectrographs like JWST/NIRSpec are therefore required to further investigate the origin of the blueshifted component, and to better understand the dynamic of the system.

\section{Summary and Conclusions}\label{sect:conclusions}
In this work, we presented high-resolution ($\sim 0\rlap{.}{\arcsec}3$) ALMA band 3 observations targeting CO(4--3) emission line and the underlying 3-mm dust continuum toward a closely-separated ($\sim 10\,{\rm kpc}$) quasar host--companion galaxy pair at $z\sim 3$. This system resides within a massive node of the Cosmic Web, hosting one of the densest concentrations of galaxies and AGN discovered so far at cosmic noon \citep{Pensabene+2024, Galbiati+2025, Travascio+2025}. 

Our accurate kinematical analysis reveals that the quasar companion (MQN01-QC) is a dynamically cold disk with a high degree of rotational support ($V_{\rm rot}/\sopra{\sigma}\approx 10$), and an estimated dynamical mass of $\simeq 2.5\times 10^{11}\,{M_{\astrosun}}$ within the inner $\simeq 4\,{\rm kpc}$. This makes MQN01-QC the most massive and fast rotating disk galaxy observed in the close proximity of a hyperluminous quasar. Although the galaxy resides in an "hostile" environment that could induce gravitational perturbations, the gas kinematics in the MQN01-QC galaxy is not deviating from that of an orderly rotating disk, and in line with the observed evolution of turbulence for disk galaxies across cosmic time. The small projected separation between the galaxies, combined with the absence of significant disturbances in the observed kinematics of the MQN01-QC galaxy disk, suggests that the quasar host may be a satellite galaxy in the early stages of a merger. The quasar host CO(4--3) emission exhibits a broad ($\sim 700\,{\rm km\,s^{-1}}$), blueshifted ($-300\,{\rm km\,s^{-1}}$) component, potentially tracing a massive molecular outflow or tidally disturbed gas caused by interaction with the massive companion galaxy.

{\rm Higher angular resolution ALMA data will be crucial to further resolve the cold gas distribution in this system and identify possible merger signatures, observations with higher sensitivity on large angular scales will be instrumental to investigate the presence of faint extended cold gaseous structures around these galaxies. Additionally, future JWST spectroscopic observations would help characterize the stellar population and ionized gas content, providing a more comprehensive view of this system. 

The MQN01 field is a unique laboratory to investigate how dense environments influence galaxy assembly and BH growth at cosmic noon. Future studies carrying out systematic comparison of the kinematical properties of galaxies across field and cluster/protocluster environments, with a careful and consistent characterization of the properties of the environments in which galaxy reside (e.g., in terms of volume, galaxy density, and overdensity definitions), will offer valuable insights into the impact of the environment on the galaxy cold gas kinematics.}

\begin{acknowledgements}
We thank Dr. Bin Ren for his valuable suggestions on optimizing the quasar PSF removal and to Prof. Massimo Dotti for his insightful guidance on galaxy dynamics. This paper makes use of the following ALMA data: ADS/JAO.ALMA\#2021.1.00793.S. ALMA is a partnership of ESO (representing its member states), NSF (USA) and NINS (Japan), together with NRC (Canada), MOST and ASIAA (Taiwan), and KASI (Republic of Korea), in cooperation with the Republic of Chile. The Joint ALMA Observatory is operated by ESO, AUI/NRAO and NAOJ. This project was supported by the European Research Council (ERC) Consolidator Grant 864361 (CosmicWeb) and by Fondazione Cariplo grant no. 2020-0902. CC has received funding from the European Union's Horizon Europe research and innovation program under grant agreement No. 101188037 (AtLAST2). This research made use of Astropy\footnote{\url{http://www.astropy.org}}, a community-developed core Python package for Astronomy \citep{AstropyI, AstropyII, AstropyIII}, NumPy \citep{Numpy}, SciPy \citep{Scipy}, Matplotlib \citep{Matplotlib}. 
\end{acknowledgements}


\bibliographystyle{aa}
\bibliography{mybib}

\begin{appendix} %First appendix
\onecolumn    
\section{Geometrical parameters of the MQN01-QC galaxy disk derived with CANNUBI}\label{app:posteriors}
\begin{figure}[!htbp]
   	\centering
   	\resizebox{0.85\hsize}{!}{
		\includegraphics{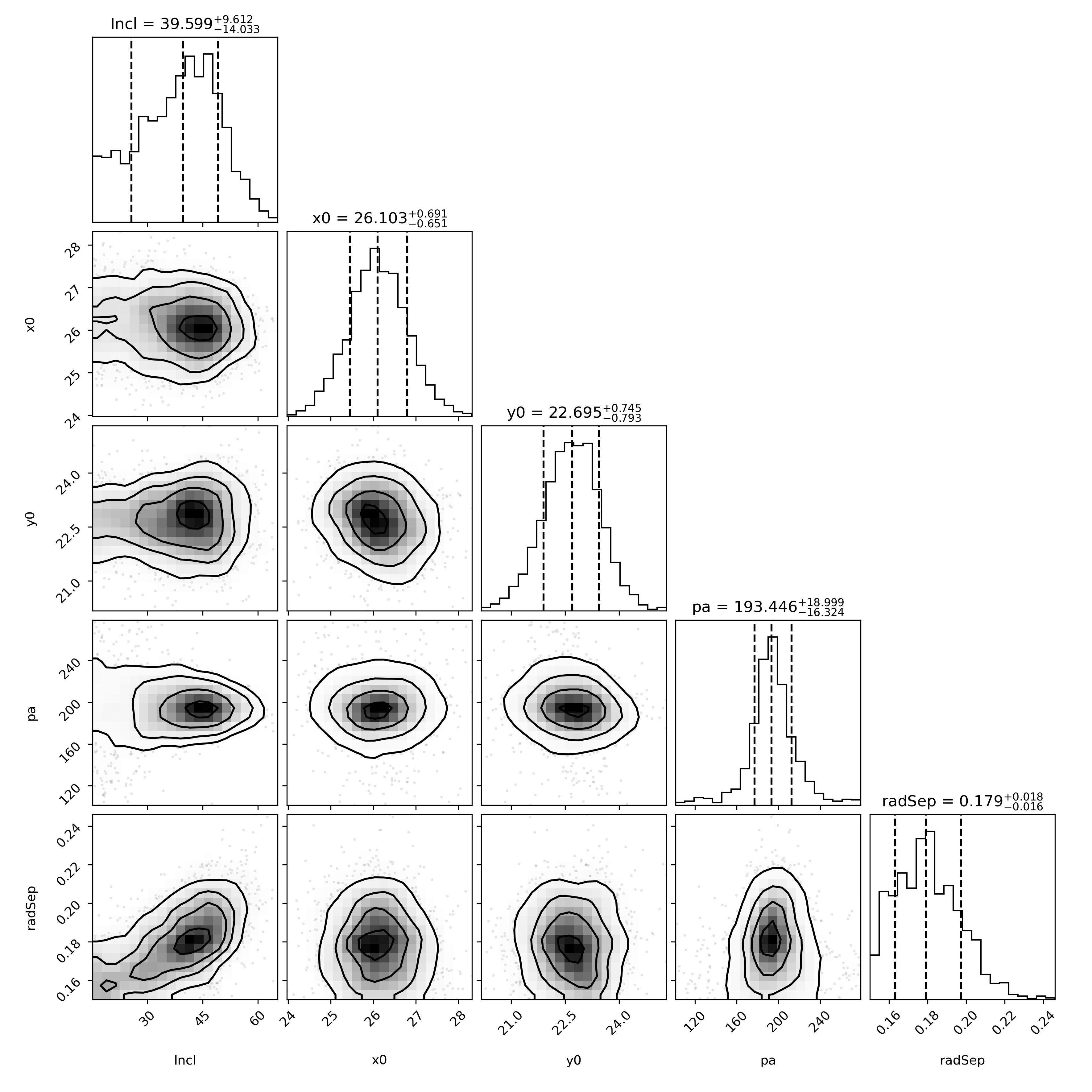}}
       \caption{Posterior probability distribution for the MQN01-QC geometrical parameters as obtained with CANNUBI. These are the disk inclination angle (incl) in degree, the disk center ($x_{0}, y_{0}$) in pixels (where 1 pix corresponds to $0\rlap{.}{\arcsec}04$), the disk position angle (pa) in degree, and the radial separation between the annuli (radSep) in arcseconds. The dashed vertical lines indicates the location of the 16th, 50th, and 85th percentiles, respectively. The best-fit values and uncertainties are derived from that and are also reported.}
       \label{fig:posterior_cannubi}
    \end{figure}

The Fig.~\ref{fig:posterior_cannubi} show the posterior probability distributions for the free parameter of the MQN01-QC disk geometry obtained with CANNUBI (see Sect.~\ref{sect:kinematics}). These include the disk inclination angle, the coordinates of the center, the position angle (pa), and the radial separation between the annulii.

\newpage
\section{Channel maps of the quasar host--MQN01-QC galaxy system}\label{app:chan-maps}
\begin{figure}[!htbp]
   	\centering
   	\resizebox{0.8\hsize}{!}{
		\includegraphics{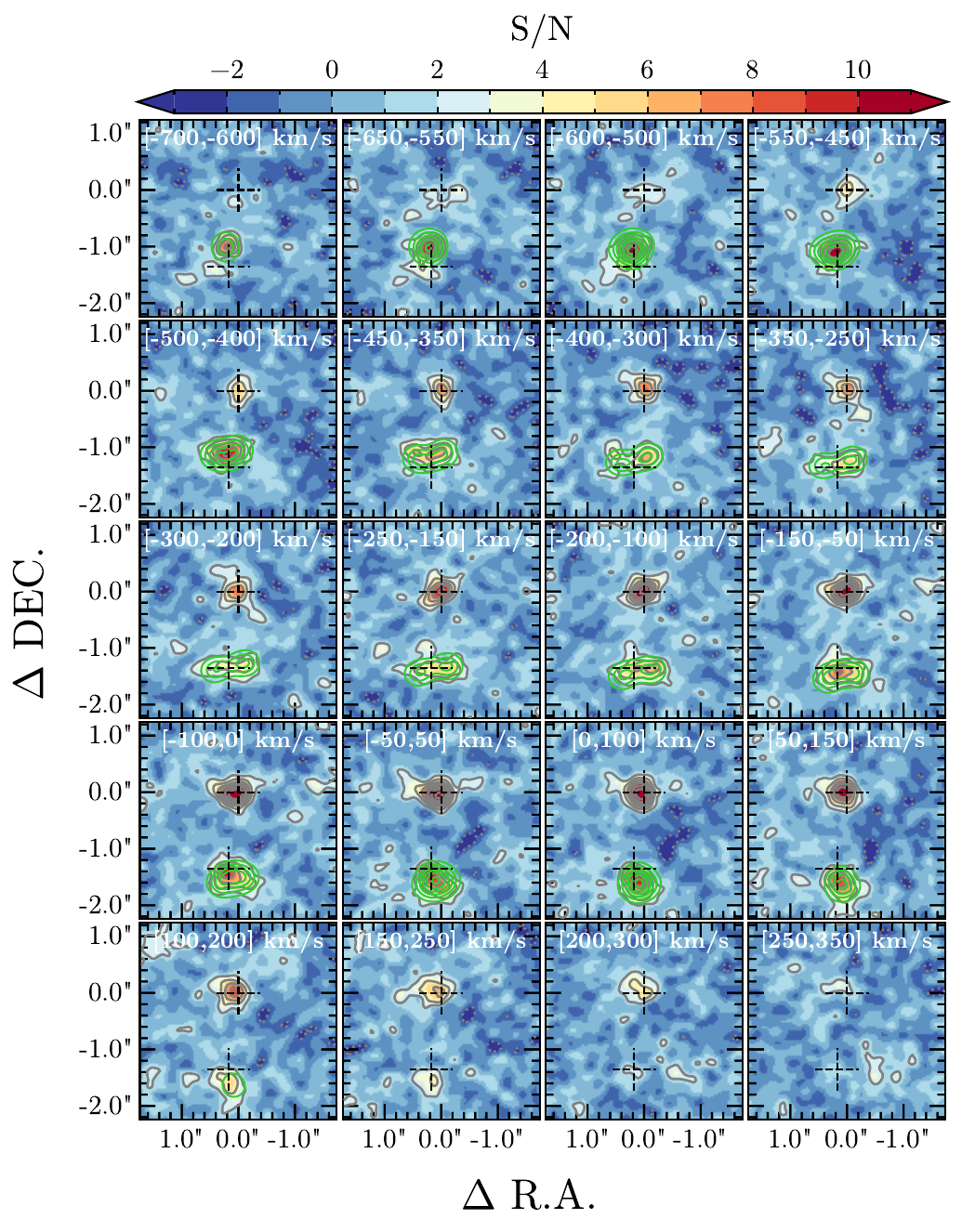}}
       \caption{Channel maps of the quasar host--MQN01-QC system extracted from the continuum-subtracted ALMA band 3 datacube. Each panel is obtained by averaging channels in the datacube within $100\,{\rm km\,s^{-1}}$-wide windows as reported at the top of each panel. The original datacube used for the kinematical modeling has a channel width of $25\,{\rm km\,s^{-1}}$. The maps are color-coded based on the signal-to-noise ratio. The gray contours show $-2\times{\rm S/N}$ (dotted), and $2n\times{\rm S/N}$ (solid) isophotes where $n\ge1$ is an integer number. The green contours show the best-fit rotating disk model as obtained with $^{\rm 3D}${\sc Barolo}.}
       \label{fig:chan_maps}
    \end{figure}

The Fig.~\ref{fig:chan_maps} shows the channel maps of the CO(4--3) line datacube of the quasar host and the MQN01-QC galaxy. We also overlay the corresponding best-fit rotating disk model of the MQN01-QC galaxy as obtained with $^{\rm 3D}${\sc Barolo}.

\end{appendix}

\end{document}